\documentclass[sigconf,9pt]{acmart}

\usepackage[english]{babel}
\usepackage{blindtext}
\usepackage{multicol}
\usepackage[normalem]{ulem}

\ccsdesc[500]{Computer systems organization~Dependable and fault-tolerant systems and networks}
\ccsdesc[500]{Computing methodologies~Distributed computing methodologies}

\keywords{Collective Communication, Distributed Systems}

\setcopyright{rightsretained}

\usepackage{tikz}
\usepackage{amsmath}
\usepackage{xspace}
\usepackage{booktabs}
\usepackage{subcaption}
\usepackage{multirow}

\usepackage{siunitx}
\usepackage{pifont}%

\newcommand{\codename}{Hoplite\xspace}

\def\Snospace~{\S{}}

\usepackage{url}

\newcommand{\autorefsuffix}[2]{\hyperref[#1]{\autoref*{#1}#2}}

\begin{document}

\copyrightyear{2021}
\acmYear{2021}
\acmConference[SIGCOMM '21]{ACM SIGCOMM 2021 Conference}{August 23--27, 2021}{Virtual Event, USA}
\acmBooktitle{ACM SIGCOMM 2021 Conference (SIGCOMM '21), August 23--27, 2021, Virtual Event, USA}\acmDOI{10.1145/3452296.3472897}
\acmISBN{978-1-4503-8383-7/21/08}

\date{}

\title{\codename: Efficient and Fault-Tolerant Collective Communication for Task-Based Distributed Systems}

\author{Siyuan Zhuang$^{\text{1}, *}$ \enskip Zhuohan Li$^{\text{1}, *}$ \enskip Danyang Zhuo$^{\text{2}}$ \enskip Stephanie Wang$^{\text{1}}$ \\ Eric Liang$^{\text{1}}$\enskip Robert Nishihara$^{\text{1}}$\enskip Philipp Moritz$^{\text{1}}$\enskip Ion Stoica$^{\text{1}}$}
\affiliation{\vspace{1mm} $^{\text{1}}$University of California, Berkeley \enskip $^{\text{2}}$Duke University\country{}}

\newcommand{\plainauthors}{Siyuan Zhuang, Zhuohan Li, Danyang Zhuo, Stephanie Wang, Eric Liang, Robert Nishihara, Philipp Moritz, Ion Stoica}
\renewcommand{\shortauthors}{Siyuan Zhuang et al.}

\begin{abstract}
Task-based distributed frameworks (e.g., Ray, Dask, Hydro) have become increasingly popular for distributed applications that contain asynchronous and dynamic workloads, including asynchronous gradient descent, reinforcement learning, and model serving. As more data-intensive applications move to run on top of task-based systems, collective communication efficiency has become an important problem. Unfortunately, traditional collective communication libraries (e.g., MPI, Horovod, NCCL) are an ill fit, because they require the communication schedule to be known before runtime and they do not provide fault tolerance.

We design and implement \codename, an efficient and fault-tolerant collective communication layer for  task-based distributed systems. Our key technique is to compute data transfer schedules on the fly and execute the schedules efficiently through fine-grained pipelining. At the same time, when a task fails, the data transfer schedule adapts quickly to allow other tasks to keep making progress. We apply \codename to a popular task-based distributed framework, Ray. We show that \codename speeds up asynchronous stochastic gradient descent, reinforcement learning, and serving an ensemble of machine learning models that are difficult to execute efficiently with traditional collective communication by up to 7.8x, 3.9x, and 3.3x, respectively.
\end{abstract}

\maketitle

\section{Introduction}
{\let\thefootnote\relax\footnote{{$^*$Equal contribution.}}}
Task-based distributed systems (e.g., Ray~\cite{moritz2018ray}, Hydro~\cite{hydro}, Dask~\cite{rocklin2015dask}, CIEL~\cite{murray2011ciel}) have become increasingly popular for developing and running distributed applications that contain asynchronous and dynamic computation and communication patterns, including asynchronous stochastic gradient descent (SGD), reinforcement learning (RL), and model serving. Today, many top technology companies have started to adopt task-based distributed frameworks for their distributed applications, such as Intel, Microsoft, Ericsson, and JP. Morgan. For example, Ant Financial uses task-based distributed systems to run their online machine learning pipeline and serve financial transactions for billions of users~\cite{keynote2020ant}.

There are two key benefits of building distributed applications on top of task-based systems. First, it is easy to express asynchronous and dynamic computation and communication patterns. A task-based system implements a \textit{dynamic task} model: a caller can dynamically invoke a task $A$, which immediately returns an \emph{object future}, i.e. a reference to the eventual return value. By passing the future as an argument, the caller can specify another task $B$ that uses the return value of $A$ even before $A$ finishes. The task-based system is responsible for scheduling \emph{workers} to execute tasks $A$ and $B$ and transferring the result of $A$ to $B$ between the corresponding workers. Second, fault tolerance is provided by the task-based system transparently. When a task fails, the task-based system quickly reconstructs the state of the failed task and resumes execution~\cite{wang2019lineagestash, sreekanti2020fault}. Well-behaving tasks do not need to roll back, so failure recovery is low cost.

As a growing number of data-intensive workloads are moving to task-based distributed systems, supporting efficient collective communication (e.g., broadcast, reduce) has become critical. Consider an RL application where the trainer process broadcasts a policy to a set of agents that use this policy to perform a series of simulations. Without the support for collective broadcast, the trainer process needs to send the same policy to every agent which causes a network bottleneck on the sender side.

Efficient collective communication is a well-understood problem in the HPC community and in distributed data-parallel training. Many collective communication libraries exist today, e.g.,  OpenMPI~\cite{graham2005open}, MPICH~\cite{mpich}, Horovod~\cite{alex2018horovod}, Gloo~\cite{gloo}, and NCCL~\cite{nccl}. However, there are two limitations of traditional collective communication implementations that make them an ill fit for dynamic task-based systems.

First, a distributed application using traditional collective communication must specify the communication pattern \emph{before} runtime. This allows the library to compute a \emph{static} and efficient data transfer schedule (e.g., ring-allreduce). For example, for \emph{synchronous} distributed data-parallel training, the application specifies that all workers participate in an allreduce communication, once per training round.

However, in task-based systems, the set of tasks or data objects participating in the collective communication is not known before runtime. One approach would be to wait until all the participating tasks and objects are ready and then compute a static data transfer schedule. Unfortunately, this design misses the opportunity to make partial progress before the entire set of participants are ready, which is critical for the performance of modern \emph{asynchronous} applications, e.g., distributed RL.

Second, because of the synchronous nature of collective communications, one process failure can cause the rest of the processes to hang. Existing solutions leave the recovery up to the application. A typical approach is to checkpoint the state of the application periodically (e.g., every hour), and when a process fails, the entire application rolls back to the previous checkpoint and restarts. Unfortunately, this can be expensive for large-scale asynchronous applications, and does not exploit the ability of tasks that are still alive in the same collective communication group as a failed task to make progress.

This raises an important question: \textit{how can we bring the efficiency of collective communication to dynamic and asynchronous task-based applications?} There are two requirements that are unique to this setting. First, the application must be allowed to specify the participants of a collective communication \emph{dynamically} (i.e., at runtime). Second, the collective communication implementation must be \emph{asynchronous}. This would allow tasks to make progress even if other tasks in the same communication group have failed. 

We design and implement \codename, an efficient and fault-tolerant collective communication layer for task-based distributed systems. \codename combines two key ideas: (1) \codename computes data transfer schedule on-the-fly as tasks and objects arrive, and \codename executes data transfer schedule efficiently using fine-grained pipelining. Collective communication can make significant progress even if only a fraction of the participants are ready. (2) \codename dynamically adapts the data transfer schedule when a failure is detected to alleviate the effects of the failed task in collective communication. This allows the live tasks to make progress. The failed task can rejoin the collective communication after being restarted and complete the communication.

We apply \codename to a popular task-based framework, Ray~\cite{moritz2018ray}. This allows us to evaluate a wide range of existing workloads on Ray. Our evaluations show that \codename can speed up an asynchronous SGD by up to 7.8x, two popular RL algorithms (IMPALA~\cite{espeholt2018impala}, and A3C~\cite{Mnih2016a3c}) on RLlib~\cite{liang2018rllib} by up to 1.9x, and 3.9x, respectively, and improve the serving throughput time of an ensemble of ML models on Ray Serve~\cite{rayserve} by up to 3.3x, with only minimal code changes and negligible additional latency in failure recovery.

This paper makes the following contributions:
\begin{itemize}
\item A distributed scheduling scheme for data transfer that provides efficient broadcast and reduce primitives for dynamic-task systems.
\item A fine-grained pipeline scheme that achieves low-latency data transfers between tasks located both on the same node or on different nodes.
\item Algorithms to adapt the schedule of the data transfers for broadcast and reduce operations which allows live tasks to make progress when other tasks that participate in the collective communication have failed, and later allow those failed tasks to rejoin.
\item We demonstrate the benefits of \codename on top of a popular task-based distributed system using several applications, including asynchronous SGD, RL, and serving an ensemble of ML models.
\end{itemize}

\section{Background}

We first describe task-based distributed systems and their benefits for developing distributed applications. We then describe the challenges of integrating efficient collective communication into them.

\begin{figure}[t!]
    \begin{subfigure}[b]{0.5\textwidth}
        \includegraphics[trim={0 0.8in 0 0},clip,width=.90\textwidth]{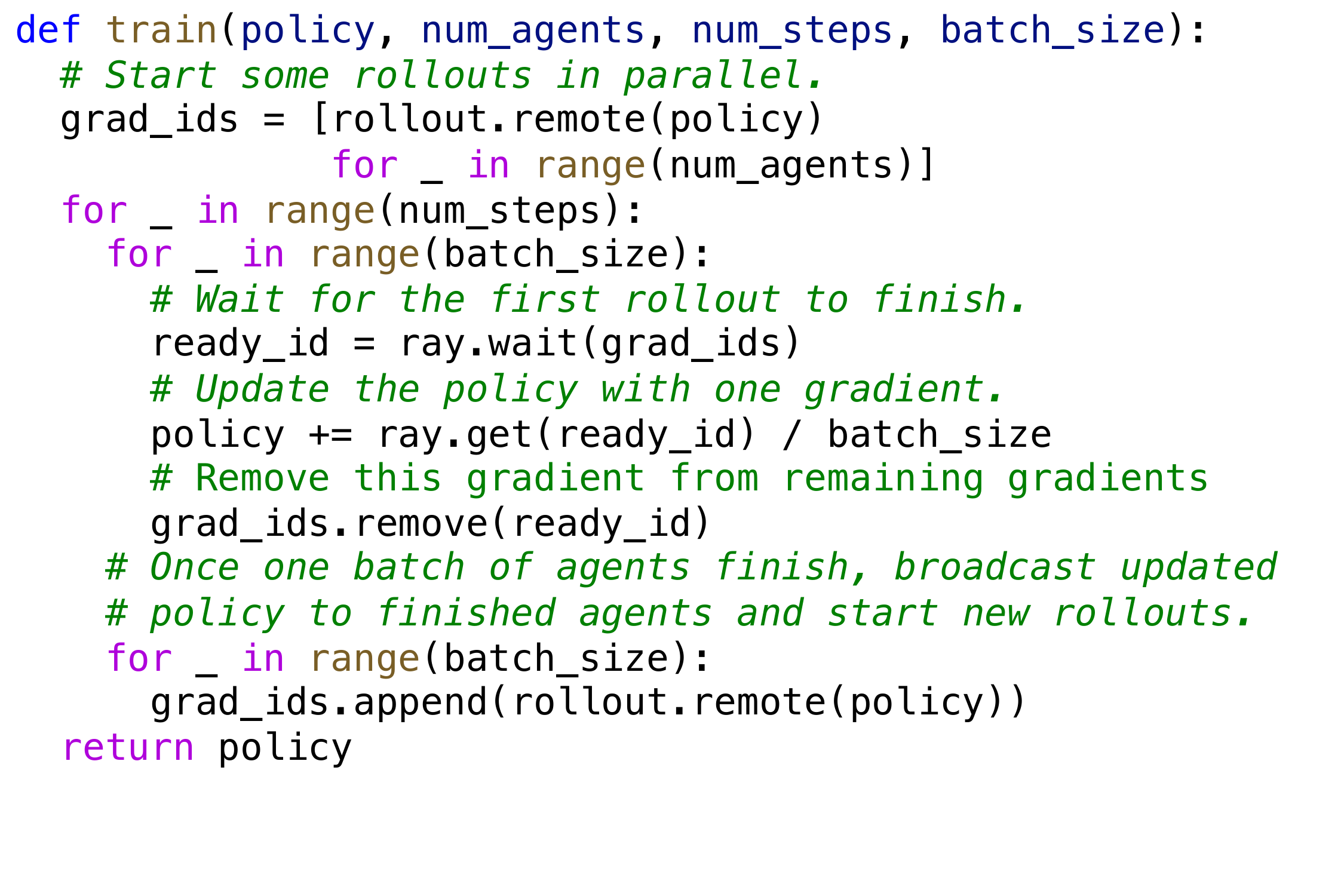}
        \subcaption{Dynamic tasks (Ray).}
        \vspace{3mm}
        \label{code:rl-example:ray}
    \end{subfigure}
    \begin{subfigure}[b]{0.5\textwidth}
        \includegraphics[trim={0 0.0in 0 0},clip,width=.90\textwidth]{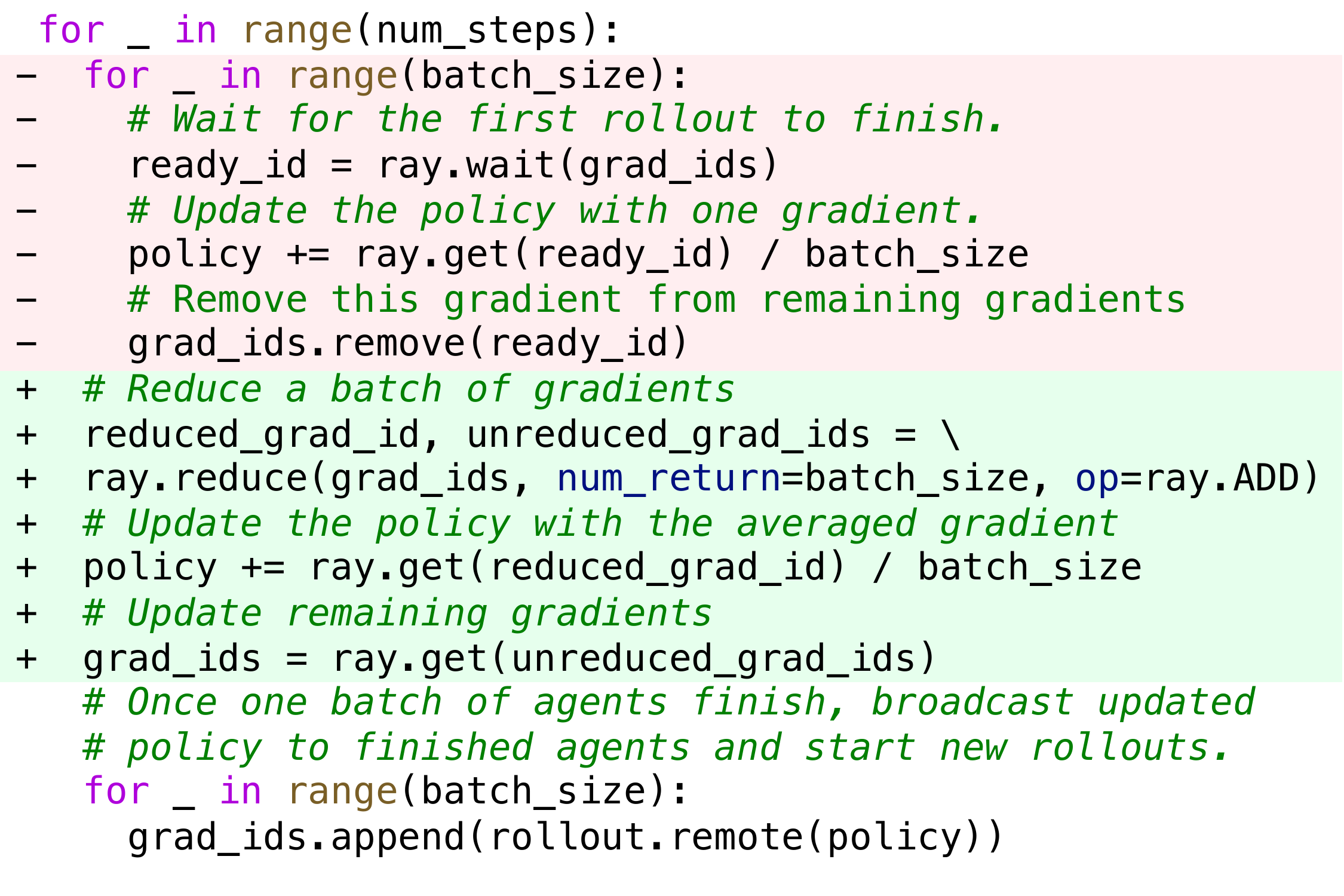}
        \subcaption{Dynamic tasks + collective comm. (Ray + Hoplite).}
        \label{code:rl-example:hoplite}
    \end{subfigure}
    \caption{
        \normalfont \small
        Pseudocode for a typical RL algorithm to learn a policy.
        \textbf{(a)} Dynamic tasks with Ray.
        Each \texttt{train} loop waits for a \emph{single} agent to finish, then asynchronously updates the current policy. The new policy is broadcast to a batch of finished agents.
        \textbf{(b)} Modifications to (a) to enable \codename.
        Each step reduces gradients from a subset of agents, updates the current policy, broadcasts the new policy.
    }
    \label{code:rl-example}
\end{figure}

\subsection{Task-Based Distributed Systems}

\begin{figure}[t]
    \begin{subfigure}[b]{0.4\textwidth}
        \includegraphics[width=\textwidth,trim=0 75 0 75, clip]{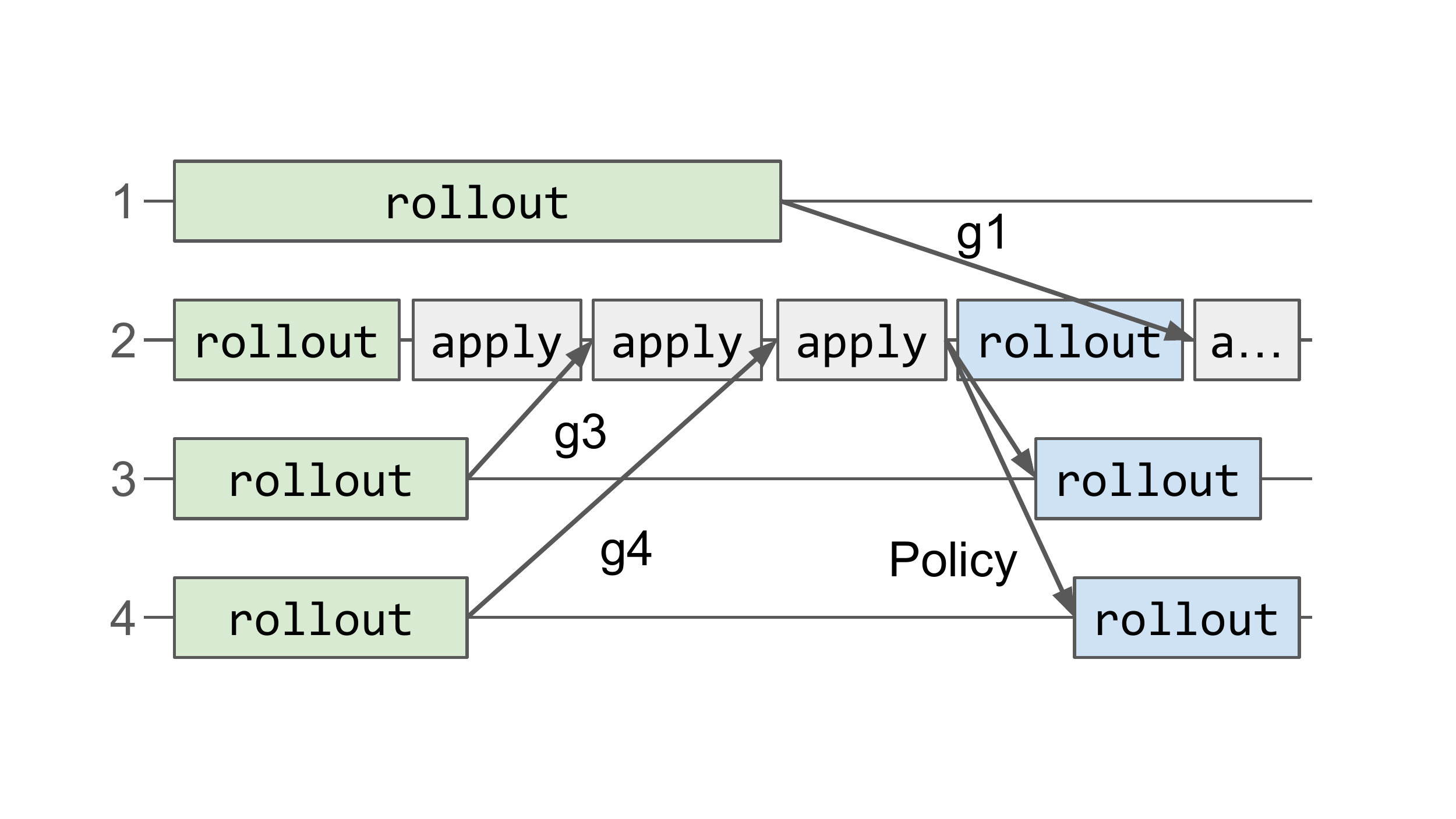}
        \subcaption{Dynamic tasks (Ray)}
        \vspace{2mm}
        \label{fig:rl-reduce:ray}
    \end{subfigure}
    \begin{subfigure}[b]{0.4\textwidth}
        \includegraphics[width=0.94\textwidth,trim=0 75 0 75, clip]{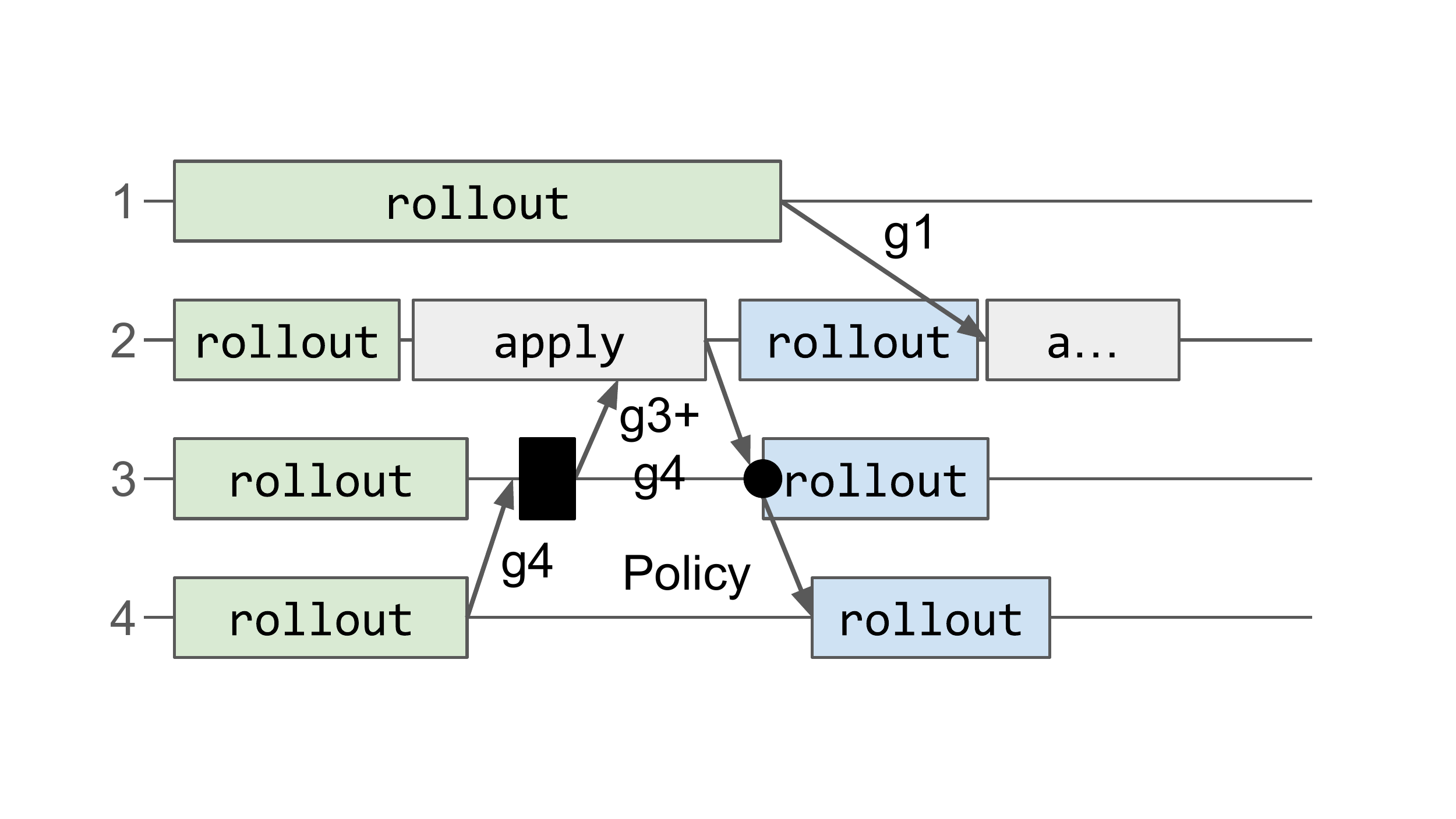}
        \subcaption{Dynamic tasks + collective comm. (Ray + Hoplite).}
        \label{fig:rl-reduce:hoplite}
    \end{subfigure}
    \caption{\normalfont \small
    Execution of a distributed RL algorithm.
    Each row is one agent.
    Boxes represent computations, and arrows represent data transfers.
    $g1$-$g4$ are the gradients produced by the agents.
    \textbf{(a)} Dynamic tasks (Ray).
    Gradients are applied immediately. A batch of three gradients is applied to the current policy before broadcasting.
    \textbf{(b)} Dynamic tasks but with efficient collective communication, in \codename.
    To reduce the network bottleneck at agent 2, agent 3 partially reduces gradients $g3$ and $g4$ (black box), and agent 3 sends the policy to agent 4 (black dot) during the broadcast.
    }
\end{figure}

The dynamic task programming model~\cite{blumofe1996cilk,murray2011ciel,moritz2018ray,rocklin2015dask,hydro} allows applications to express asynchronous and dynamic computation and communication patterns. For instance, \autoref{code:rl-example:ray} shows how to implement an \emph{asynchronous} RL algorithm that updates the policy with agent results one at a time, choosing them \emph{dynamically} based on the order of availability. Once a batch of agent results have been applied, the resulting policy is sent to each finished agent to begin the next round of rollout. This allows an agent that has a fast rollout not need to wait for a worker that has a slow rollout~(\autoref{fig:rl-reduce:ray}). Today, most RL algorithms~\cite{espeholt2018impala, Mnih2016a3c} leverage this type of asynchronous execution for efficient training.

To support this type of asynchronous communication, task-based distributed systems rely on a \textit{distributed object store} to transfer objects between tasks. The object store consists of a set of \emph{nodes}, each of which buffers a (possibly overlapping) set of application objects. Each node serves multiple workers, which can read and write directly to objects in its local node via shared memory. A sender task stores the output into the object store and exits, allowing it to release critical resources (e.g., CPU, GPU, memory) before the receiver tasks are even scheduled. When receiver tasks are ready, they directly fetch the object from the distributed object store. As is standard~\cite{murray2011ciel,moritz2018ray}, the object store enforces object immutability and uses a distributed object directory service to map each object to its set of node locations. In addition, task-based distributed systems support fast failure recovery~\cite{wang2019lineagestash, sreekanti2020fault} by reconstructing the failed task. Well-behaving tasks do not roll back to keep recovery low cost.

However, if the gradients and the model are large enough in the above RL example, task-based distributed systems incur significant overheads from inefficient communication. For example, the trainer (agent 2) in \autoref{fig:rl-reduce:ray} can become a network throughput bottleneck since it has to receive the gradient and also send the new policy from/to each agent individually. This bottleneck becomes more severe when the number of agents increases.

\subsection{Challenges in Collective Communication}

Efficient collective communication has well-known solutions in HPC community and in distributed data-parallel SGD. Many traditional collective communication libraries exist, including Gloo~\cite{gloo}, Horovord~\cite{alex2018horovod}, OpenMPI~\cite{graham2005open}, MPICH~\cite{mpich}, and NCCL~\cite{nccl}. They can use efficient data transfer schedule (e.g., ring-allreduce, tree-broadcast) to mitigate communication bottlenecks in distributed applications.

There are two application requirements for using traditional collective communication libraries. First, the communication pattern has to be statically defined before runtime. This is easy for applications that have a bulk-synchronous parallel model. For example, in synchronous data-parallel SGD, all the workers compute on their partitioned set of training data and synchronize the model parameters using allreduce. Second, when any worker fails, all the workers participating in the collective communication hang, and applications are responsible for fault tolerance. For HPC applications, this is typically solved by checkpointing the entire application periodically (e.g., per-hour), and when a process fails, the entire application rolls back to a checkpoint and re-execute.

Unfortunately, these two assumptions are fundamentally incompatible with task-based distributed systems. First, tasks are dynamically invoked by the task-based system's scheduler. This means it is possible that, when collective communication is triggered, only a fraction of the participating tasks are scheduled. For example, on existing task systems, broadcast is implicit: a set of tasks fetch the same object. When only a subset of the receivers are scheduled, it is not possible to build a static broadcast tree without knowing how many total receivers and where and when the receivers will be scheduled. \textit{Therefore, a collective communication layer for a task-based system should adjust data transfer schedule at runtime based on task and object arrivals.}

Second, fast failure recovery is an important design goal for task-based system~\cite{wang2019lineagestash, sreekanti2020fault}, because many asynchronous workloads have tight SLO requirement (e.g., model serving). In existing task systems, this is done by reconstructing and re-executing failed tasks only. If traditional collective communication libraries are used, a failed task causes the rest of the participating tasks to hang. Thus, \textit{a collective communication layer for task-based systems has to be fault-tolerant: allowing well-behaving tasks to make progress when a task fails and allowing the failed task to rejoin the collective communication after recovery.}

\section{Design}
\label{sec:design}
\codename is an efficient and fault-tolerant collective communication layer for task-based distributed systems. At a high level, \codename uses two techniques: (1) decentralized fault-tolerant coordination of data transfer for reduce and broadcast, and (2) pipelining of object transfers both across nodes and between tasks and the object store. 

We first present a send-receive example workflow using \codename's core API (\autoref{tab:api}). We then describe \codename's object directory service, pipelining mechanism to reduce latency, and fault-tolerant receiver-driven coordination scheme for efficient object transfer in details.

\subsection{\codename's Workflow}
\label{sec:overview}

\begin{figure}[t]
    \centering
    \includegraphics[trim={0 5.0in 2.0in 0},clip,width=.45\textwidth]{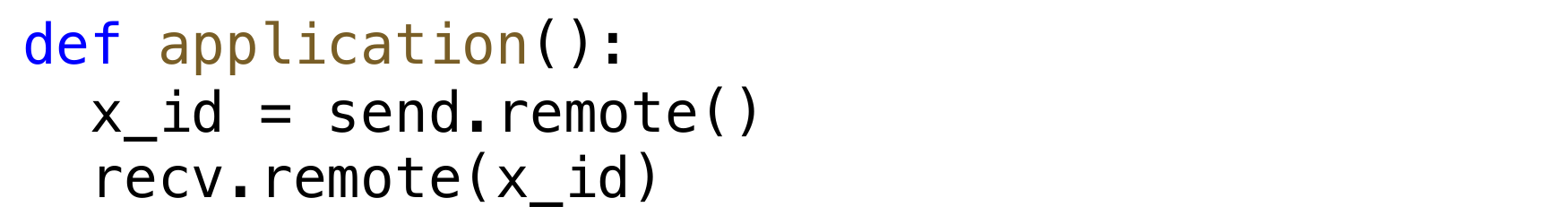}
    \includegraphics[width=0.45\textwidth,trim=95 90 100 70, clip]{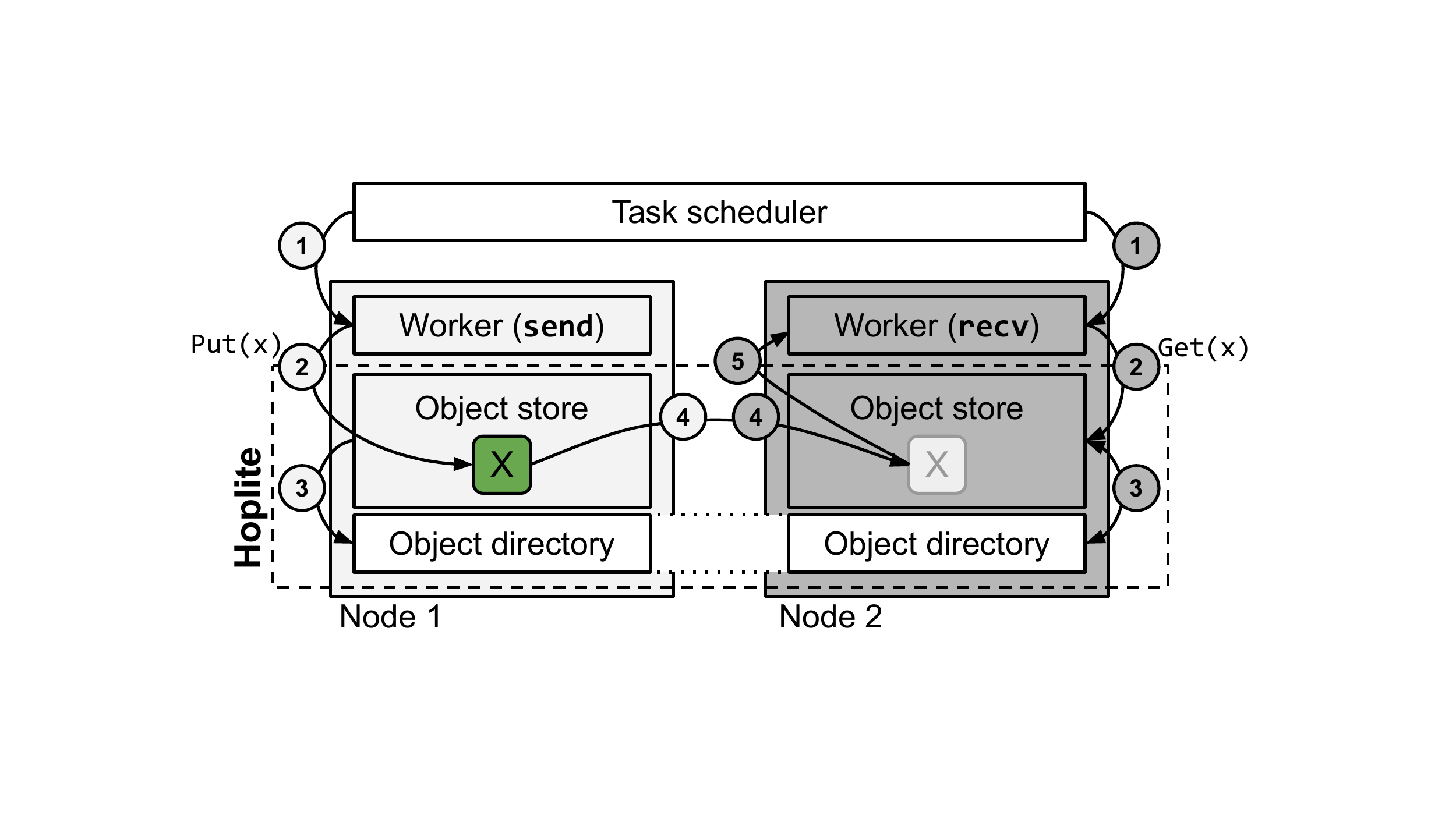}
    \caption{\normalfont \small
    Example of a send and receive dynamic task program on a 2-node cluster (\texttt{N1} and \texttt{N2}).
    The task-based system consists of a pool of workers per physical node and a scheduler.
    \codename consists of one local object store per node and a global object directory service, which is distributed across physical nodes.
    }
    \label{fig:arch}
\end{figure}

\begin{table*}[ht]
\centering

\begin{tabular}{l l}
  \toprule
  \textbf{Core Interfaces:} & \textbf{Description} \\
  Buffer buffer$\leftarrow$ Get(ObjectID object\_id) & Get an object buffer from an object id.\\
  Put(ObjectID object\_id, Buffer buffer) & Create an object with a given object id and an object buffer. \\
  Delete(ObjectID object\_id) & Delete all copies of an object with a given object id.\\
                              & \hspace{0.5cm} Called by the task framework once an object is no longer in use. \\
  Reduce(ObjectID target\_object\_id, int num\_objects, & Create a new object with a given object id from a set of objects \\
  \hspace{0.5cm} \{ObjectID source\_object\_id, ...\}, ReduceOp op) & \hspace{0.5cm} using a reduce operation (e.g, sum, min, max). \\
  \bottomrule
\end{tabular}
\caption{
    \normalfont \small
    Core \codename APIs.
    The application generates an \texttt{ObjectID} with a unique string and can pass an \texttt{ObjectID} by sending the string.
}
\label{tab:api}
\end{table*}

Our example creates a \texttt{send} task that returns \texttt{x\_id} (a future), which is then passed into a \texttt{recv} task. In \codename, we use an \texttt{ObjectID} to represent a future or a reference to an object. During execution, the application first submits the tasks to the task scheduler. The scheduler then chooses a worker to execute each task (step 1, \autoref{fig:arch}), e.g., based on resource availability. According to the application, \texttt{recv} cannot start executing until it has the value returned by \texttt{send}. Note that the task-based system does not require the scheduler to schedule tasks in a particular location or order, i.e. the \texttt{recv} task may be scheduled \emph{before} \texttt{send}.

In step 2, the task workers call into \codename to store and retrieve objects. On node 1, the \texttt{send} worker returns an object with the unique ID \texttt{x\_id}. This object must be stored until the \texttt{recv} worker has received it. Thus, the \texttt{send} worker calls \texttt{Put(x)} on \codename, which copies the object from the worker into the local object store (step 2 on \texttt{N1}, \autoref{fig:arch}). This frees the worker to execute another task, but incurs an additional memory copy between processes to store objects.

Meanwhile, on node 2, the \texttt{recv} worker must retrieve the object returned by \texttt{send}. To do this, it calls \texttt{Get(x)} on \codename, which blocks until the requested object has been copied into the worker's local memory (step 2 on \texttt{N2}, \autoref{fig:arch}). In step 3, \codename uses the \emph{object directory service} to discover object locations and coordinate data transfer, in order to fulfill the client's \texttt{Put} and \texttt{Get} requests. In the example, the \codename object store on node 1 publishes the new location for the object \texttt{x} to the directory (step 3 on \texttt{N1}, \autoref{fig:arch}). Meanwhile, on node 2, the \codename object store queries the directory for a location for \texttt{x} (step 3 on \texttt{N2}, \autoref{fig:arch}).

\codename's object directory service (\autoref{sec:directoryservice}) is implemented as a sharded hash table that is distributed throughout the cluster (\autoref{fig:arch}). Each shard maps an \texttt{ObjectID} to the current set of node locations. When there are multiple locations for an object, the directory service can choose a \emph{single} location to return to the client. The object store also maintains information about objects that have only been partially created to facilitate object transfer pipelining~(\autoref{sec:pipelining}). For example, in \autoref{fig:arch}, the object store on node 1 publishes its location to the object directory as soon as \texttt{Put(x)} is called, even if the object hasn't been fully copied into the store yet. This allows node 1 to begin sending the object to node 2 while it is still being copied from the \texttt{send} worker.

Finally, in step 4, the \codename object store nodes execute the data transfer schedule specified by the object directory's reply to node 2. Node 1 is the only location for \texttt{x}, so node 2 requests and receives a copy from node 1 (step 4). Node 2 then copies the object from its local store to the \texttt{recv} worker (step 5 in \autoref{fig:arch}), which again can be pipelined with the copy over the network.

\codename provides two efficient collective communication schemes. \codename implements efficient broadcast through coordination between the object directory service and the workers (\autoref{sec:coordination}), without an explicit primitive. For reduce, \codename exposes an explicit \texttt{Reduce} call to the task-based system. It is necessary because this lets \codename know that these objects are indeed reducible (i.e., the operation is commutative and associative). Because an \texttt{ObjectID} is a future that the object value may not be ready yet, the \texttt{Reduce} call also has a \texttt{num\_objects} input in case the user wants to reduce a subset of the objects, giving \codename the flexibility to choose which \texttt{num\_objects} objects to reduce given their arrival time in the future. \autoref{code:rl-example:hoplite} shows how to modify the RL example to use \codename. This allows the trainer to aggregate gradients from a dynamic set of agents efficiently (\autoref{fig:rl-reduce:hoplite}).

Whenever a task fails, \codename recomputes a data transfer schedule to avoid using the failed task in the collective communication, and all the rest of the tasks can keep making progress (\autoref{sec:fault-tolerance}). \codename does not change how task-based distributed system tolerate failures. The underlying task-based distributed system can quickly reconstruct the state of the failed task using their built-in mechanism~\cite{wang2019lineagestash}. Once the state of the task is reconstructed, the task resumes.

\subsection{Object Directory Service}
\label{sec:directoryservice}

The object directory service maintains two fields for each object: (1) the size of the object, and (2) the location information. The location information is a list of node IP addresses and the current progress of the object on that node. We use a single bit to represent the object's progress: either the node contains a partial or a complete object. We store both so that partial object copies can immediately act as senders, for both broadcast and reduce~(\autoref{sec:coordination}).

\codename's directory service supports both synchronous and asynchronous location queries. Synchronous location queries block until corresponding objects are created and locations are known. Asynchronous location queries return immediately, and the object directory service publishes any future locations of the object to the client.

A node writes object locations to the object directory service in two conditions: when a local client creates an object via \texttt{Put} and when an object is copied from a remote node. In each case, the node notifies the object directory service twice: once when an object is about to be created in the local store and once when the complete object is ready. We differentiate between partial and complete objects so that object store nodes with complete copies can be favored during a broadcast or reduce~(\autoref{sec:coordination}). 

\emph{Optimization for small objects.}
Querying object location can introduce an excessive latency penalty for fetching small objects, and the overhead of computing efficient object transfer schedule is usually not worthwhile for small objects in our use cases. Therefore, we implement a fast path in the object directory service. For small objects (<64KB), we simply cache them in the object directory service, and when a node queries for their location, the object directory service directly returns the object buffers. Similar to object in the per-node stores, cached objects must be freed by the application via the \texttt{Delete} call when no longer in use.

\subsection{Pipelining}
\label{sec:pipelining}
\codename uses pipelining to achieve low-latency transfer between processes and across nodes for large objects. This is implemented by enabling a receiver node to fetch an object that is incomplete in a source node. An object can be incomplete if the operation that created the object, either a \texttt{Put} from the client or a copy between object store nodes, is still in progress. To enable fetching incomplete objects, as shown in the previous section (\autoref{sec:directoryservice}), the object directory service also maintains locations of incomplete copies. Then, when an object store receives a \texttt{Get} operation, it can choose to request the object from a store with an incomplete copy.

By pipelining data transfers across nodes using the object directory service as an intermediary, it becomes simple to also pipeline higher-level collective communication primitives, such as a reduce followed by a broadcast~(\autoref{fig:rl-reduce:hoplite}). Within the reduce, a node can compute a reduce of a subset of the input objects and simultaneously send the intermediate result to a downstream node. The downstream node can then compute the final reduce result by computing on the intermediate result as it is received and simultaneously send the final result to any broadcast receivers that have been scheduled. A broadcast receiver can then also simultaneously send the final result to any other broadcast receivers.

Piplining between the task worker and local store on the same node is also important to hide \texttt{Put} and \texttt{Get} latency for large objects (steps 2 and 5 in \autoref{fig:arch}). The reason is that using the distributed object store requires two additional data copies other than the minimum needed to transfer data over the network. The sender task worker must copy to its local store, and then the receiving local store must also copy to its local worker. Our observation is that the additional memory copy latency can be masked by the network transfer if the memory copy is asynchronous. When a sender task calls \texttt{Put}, \codename immediately notifies the object directory service that the object is ready to transfer. A receiver can then fetch the object before the entire object is copied into the sender node's local store. The receiver side's pipelining mechanism is similar. When the receiver task calls \texttt{Get}, the receiver task starts to copy the object from the local store before the local store has a complete object.

By combining cross-node and in-node pipelining, \codename enables end-to-end object streaming between the sender and receiver tasks, even when there are multiple rounds of collective communication in between.

\emph{Optimization for immutable get.}
Although \codename objects are immutable, the receiver task still copies the object data from its local store during a \texttt{Get}, in case it modifies the buffer later on. However, if it only needs read access to the object, then \codename can directly return a pointer inside the local store. Read-only access can be enforced through the front-end programming language, e.g., with \texttt{const} in C++.

\subsection{Receiver-Driven Collective Communication}
\label{sec:coordination}
\codename's receiver-driven coordination scheme optimizes data transfer using distributed protocols. In \codename, data transfer happens in two scenarios: either a task calls \texttt{Get} to retrieve an object with a given \texttt{ObjectID}, or a task calls \texttt{Reduce} to create a new object by reducing a set of other objects with a reduce operation (e.g., sum, min, max).

\subsubsection{Broadcast}
Broadcast in a task-based distributed system happens when a group of tasks located on multiple nodes want to get the same object from its creator task. Specifically, a \emph{sender} task from node \texttt{S} creates an object with \texttt{Put} and a group of \emph{receiver} tasks \texttt{R1}, \texttt{R2}, ... fetches it using \texttt{Get}. For the receiver tasks that locate on different nodes from the sender task, their corresponding receiver nodes will fetch the object from sender node's local object store to the receiver nodes' local object store. To simplify the description of our method, we assume that the sender task and the receiver tasks locate on different nodes and use the sender \texttt{S} and the receiver \texttt{R1}, \texttt{R2}, ... to also refer to the local object store on the nodes.

Broadcast in a task-based distributed system is challenging because we have no knowledge of the tasks, including where these tasks are located and when these tasks fetch the object. If all receivers simply fetch the object from the sender, the performance will be restricted by the sender's upstream bandwidth. Traditional collective communication libraries can generate a static tree where the root is the sender node to mitigate the throughput bottleneck. The goal of \codename's receiver-driven coordination scheme is to achieve a similar effect but using decentralized protocols. Inspired by application-level broadcast~\cite{castro2002multicast,castro2003splitstream} in peer-to-peer systems that uses high-capacity nodes to serve as intermediate nodes in the broadcast tree, we use receivers who receives the object earlier than the rest as intermediates to construct a broadcast tree.

\begin{figure}[t]
\centering
\includegraphics[width=0.4\textwidth]{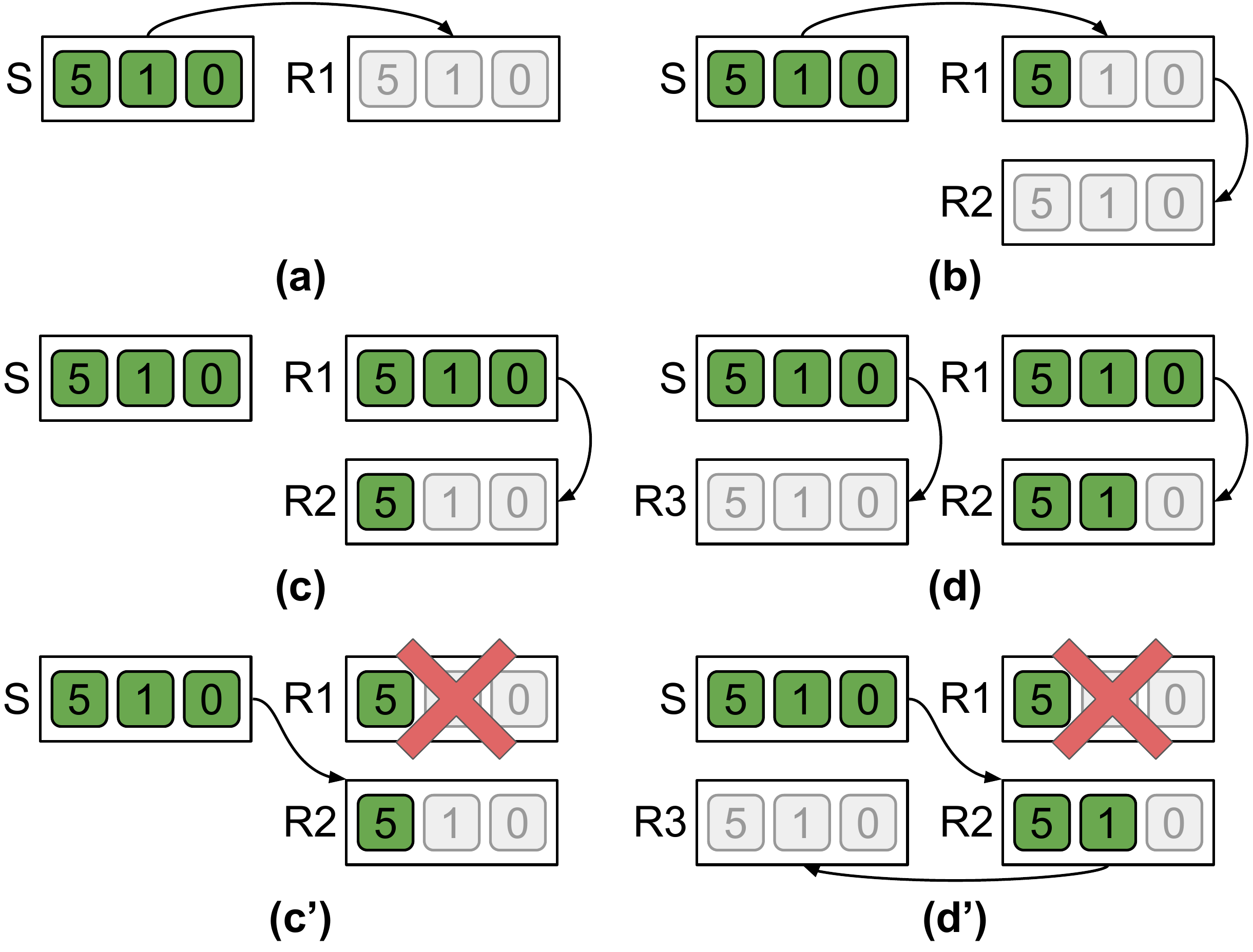}
\caption{
\normalfont \small
An example of broadcasting an object (integer array \{5, 1, 0\}) from a sender (\texttt{S}) in \codename, when the receivers (\texttt{R1}-\texttt{R3}) arrive at different times. \textbf{(a)} - \textbf{(d)} show the broadcast process without failure. \textbf{(c')} and \textbf{(d')} show the broadcast process when \texttt{R1} fails after \textbf{(b)}.
}
\label{fig:multicast}
\end{figure}

When a receiver \texttt{R} wants to fetch a remote object, it first checks if the object is locally available, or there is an on-going request for the object locally. If so, the receiver just waits until it gets the completed object. This avoids creating cyclic object dependencies. Otherwise, \texttt{R} queries the object directory service for the object's location. The object directory service first tries to return \emph{one} location with a complete copy. If none exist, then the object directory service returns one of the locations holding a partial copy.
This is so that partial objects can also act as intermediate senders, but locations with complete copies are favored.

When the location query replies, \texttt{R} also removes the location returned from the directory and immediately add itself to the object directory as a location with a partial copy to enable pipelining.
Once the data transfer is complete, the receiver adds the sender's location back to the object directory service and mark itself as a location with a complete copy. This makes sure that, for each object, a node can only send to one receiver at a time, thus mitigating bottlenecks at any single node.

\autoref{fig:multicast} shows an example of a broadcast scenario in \codename. In \autorefsuffix{fig:multicast}{a}, the first receiver \texttt{R1} starts to fetch the object from the sender \texttt{S}.
In \autorefsuffix{fig:multicast}{b}, \texttt{S} is still sending to \texttt{R1}, so it does not appear in the object directory when the second receiver \texttt{R2} arrives. Thus, \texttt{R2} fetches the object from \texttt{R1}, the partial copy.
In \autorefsuffix{fig:multicast}{c}, \texttt{R1} has finished receiving, but is still sending to \texttt{R2}.
Then, the object directory contains \texttt{S} and \texttt{R2} as a complete and partial location, respectively.
In \autorefsuffix{fig:multicast}{d}, \texttt{R3} queries the object directory, which chooses \texttt{S} over \texttt{R2} as the sender because \texttt{S} has a complete object.

\begin{figure*}[t]
    \begin{subfigure}[b]{0.45\textwidth}
        \centering
        \includegraphics[width=.85\textwidth]{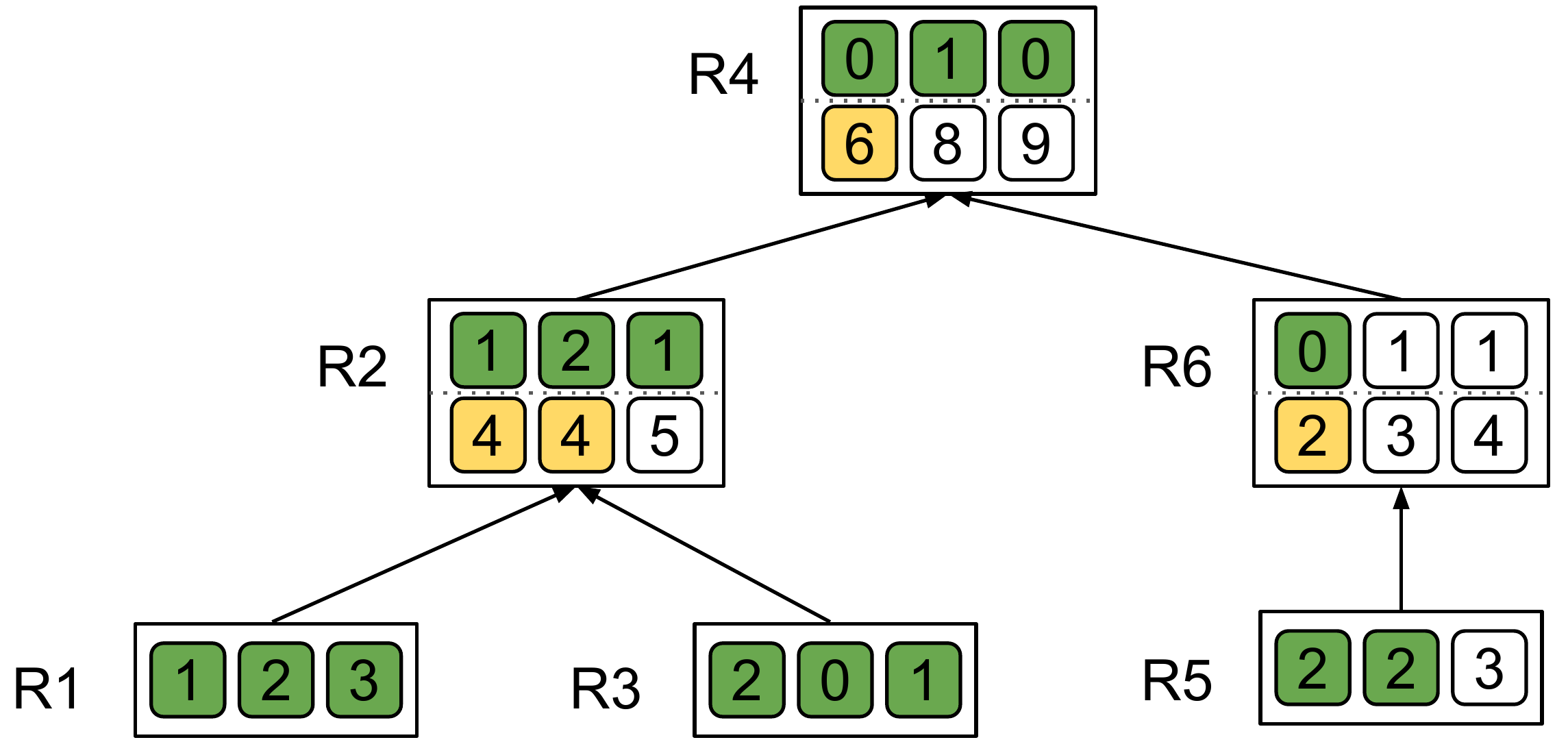}
        \subcaption{Reduce Tree.}
        \label{fig:reduce-tree}
    \end{subfigure}
    \begin{subfigure}[b]{0.45\textwidth}
        \centering
        \includegraphics[width=.85\textwidth]{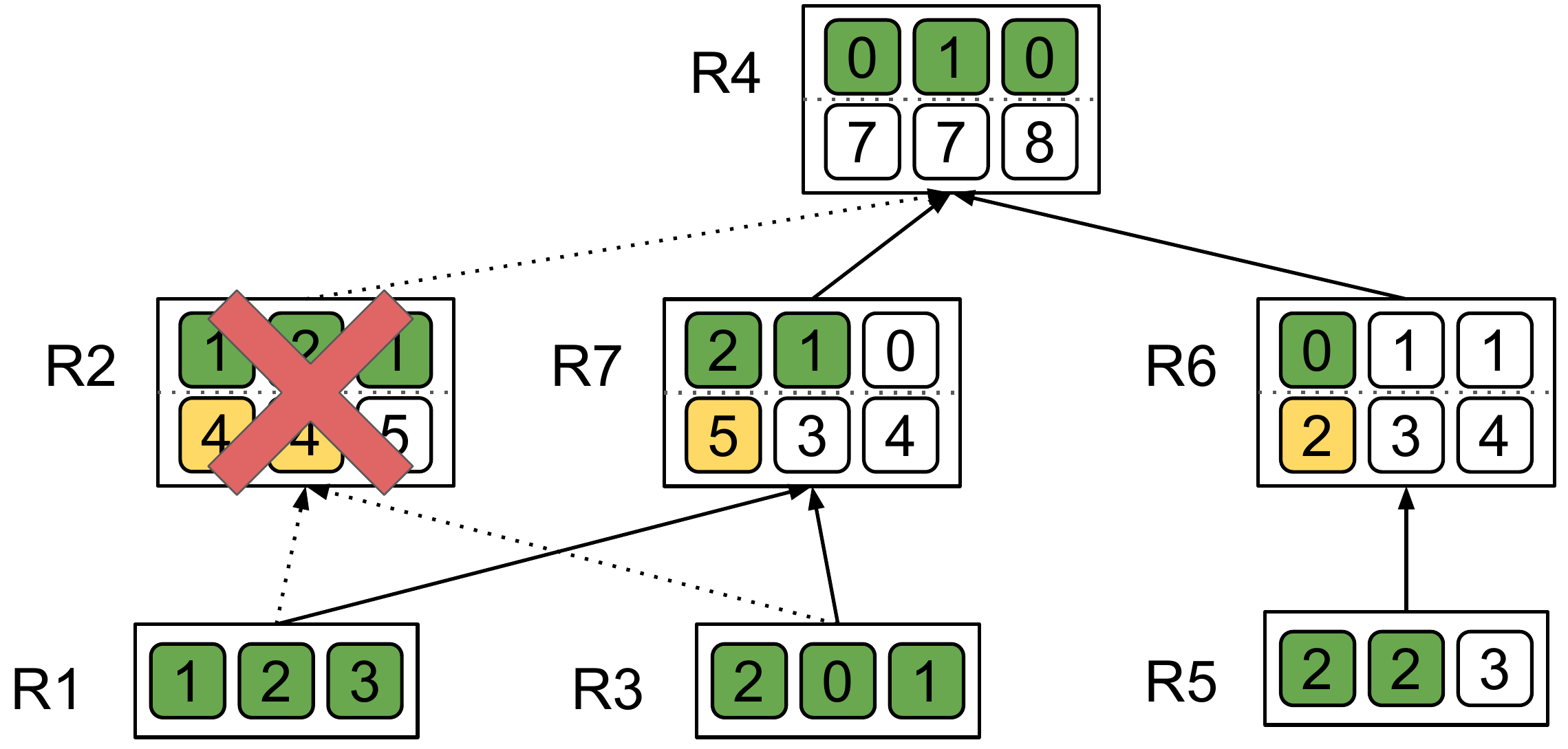}
        \subcaption{Reduce Tree with failure.}
        \label{fig:reduce-tree-faulttolerant}
    \end{subfigure}
    \caption{
        \normalfont \small Examples of reduce where the objects arrive in the order of \texttt{R1}, \texttt{R2}, ..., \texttt{R6}. The numbers on the top of each node (and the numbers in leaf nodes) represent the object to reduce and green blocks means the fraction of the object that is ready. The numbers on the bottom of each node represent the reduced result and yellow blocks means the fraction of the object that has been reduced. Each intermediate node is responsible to reduce the subtree rooted at it.
        \textbf{(a)} An example reduce tree consists of 6 objects. \textbf{(b)} The reconstructed reduce tree after \texttt{R2} fails.
    }
    \label{fig:reduce}
\end{figure*}

\subsubsection{Reduce}
Reduce happens when a task in a task-based distributed system wants to get a reduced object (e.g., summed or maximal object) from a list of objects. In \codename, this happens via a \texttt{Reduce} call. Similar to broadcast, we assume that each object to reduce is located on a separate node and we use \texttt{R1}, \texttt{R2}, ... to represent both the object and the local object store on a node that stores the corresponding object. Note that in a task-based distributed system, the objects to reduce can become ready to reduce in any arbitrary order.

How to reduce objects efficiently to accommodate dynamic object creation is more challenging than broadcast. Broadcast is simpler because a receiver can fetch the object from any sender, and \codename thus has more flexibility to adapting data transfer schedule. For reduce, we need to make sure all the objects are reduced once and only once: when one object is added into a partial reduce result, the object should not be added into any other partial results. 

In \codename, we choose to use a tree-structured reduce algorithm, while the question is what type of tree to use. Let's think about reducing $n$ objects. Without the support of collective communication in task-based distributed systems, each node sends the object to a single receiver. Let's assume that the network latency is $L$ , network bandwidth is $B$, and the object size is $S$. This approach's total reduce running time is $L + \frac{nS}{B}$. The $L$ term is due to the network latency, and $\frac{nS}{B}$ is due to the receiver's bandwidth constraint, because every node has to send the object in to it. This is a special kind of tree where the degree of the root is $n$. When object size is very small (i.e., $\frac{S}{B}$ is negligible), the performance of this kind of tree is the best. 

To mitigate the bandwidth bottleneck at the receiver, we can generalize this $n$-nary tree to a $d$-nary tree. When we use a $d$-nary tree, the total running time is $L\log_d n + \frac{dS}{B}$. It reduces the latency due to the bandwidth constraint but incurs additional latency because the height of the tree grows to $\log_d n$. If an object is very large (i.e., $\frac{S}{B} \gg L$), we can set $d = 1$. This means all the nodes are in a single chain, and its running time is $nL + \frac{S}{B}$. Note that we only need to incur $\frac{S}{B}$ for transferring the actual content of the object, because we use fine-grained pipelining, i.e., intermediate nodes send the partially reduced object to the next node. As we can see here, the optimal choice of $d$ depends on the network characteristics, the size of the object, and the number of participants (objects). In other words, we choose the $d$ to minimize the total latency:
\begin{equation}
    T(d) = \begin{cases}
    nL + \frac{S}{B} & \text{if } d=1; \\
    L\log_dn + \frac{dS}{B} & \text{otherwise.} 
    \end{cases}
\end{equation}
During runtime, \codename will automatically chooses the optimal $d$ based on an empirical measure of these three factors. 

Once the topology of the tree is determined, we need to assign nodes into the tree. Here we want to allow \texttt{Reduce} to make significant progress even with a subset of objects. To do so, we assign arriving objects with a generalized version of in-order tree traversal. For a $d$-nary tree, for each node, we traverse the first child, the node itself, the second child, third child, ..., and the $d$-th child. \autoref{fig:reduce-tree} shows an example for reducing 6 objects with a binary tree. Note that though MPI also supports tree-reduce, our method is completely different: MPI's tree is constructed statically, and our tree is constructed dynamically taking the object arrival sequence into consideration.

If a task only wants to reduce a subset of objects (i.e., \texttt{num\_object} is smaller than the size of the source object list in \texttt{Reduce}), the tree construction process stops when there are \texttt{num\_object} objects in the tree. For example, if the task wants to reduce 6 out of 10 objects, then the earliest arriving 6 objects are in the reduce tree structured as \autoref{fig:reduce-tree}.

An application can also specify the inputs of a \texttt{Reduce} incrementally, i.e. by passing the \texttt{ObjectID} result of one \texttt{Reduce} operation as an input of a subsequent \texttt{Reduce} operation. The data transfer for composed \texttt{Reduce} operations will naturally compose together. In particular, as soon as the first \texttt{Reduce} output is partially ready, it will be added to the object directory service, where it will be discovered by the downstream \texttt{Reduce} coordinator. The first output can then be streamed into the downstream \texttt{Reduce}. 

\subsubsection{AllReduce} AllReduce is a synchronous collective communication operation that is useful for synchronous data-parallel training. Optimizing allreduce is not our design goal: people usually do synchronous data-parallel training on specialized distributed systems that are optimized for bulk-synchronous workloads (e.g., TensorFlow~\cite{abadi2016tensorflow}, PyTorch~\cite{paszke2019pytorch}) rather than on task-based distributed systems. In \codename, a developer can express allreduce by concatenating reduce and broadcast.

\subsection{Fault-Tolerant Collective Communication}
\label{sec:fault-tolerance}
In the previous subsection, we assume that there is no task failures. However, task failures can happen in a task-based distributed systems for various reasons, including (1) the node that the task is running on crashes, (2) the node runs out of available memory and has to kill the task, and (3) the task encounters a runtime error. Task-based distributed systems already support transparent fault-tolerance to tasks~\cite{wang2019lineagestash}, but adding collective communication support requires us to dynamically change data transfer schedule when a fraction of the tasks fail when participating in the collective communication. This is because we do not want a failed task to block collective communication, and we want to allow a recovered task to rejoin an existing collective communication.

\subsubsection{Broadcast} When a sender failure is detected by the receiver in broadcast, the receiver immediately locate another sender by querying the object directory again. The new sender only needs to send the remaining object that the receiver does not have. A failed task can rejoin broadcast transparently because the failed task can simply call \texttt{Get} on the same \texttt{ObjectID} to fetch the object. Implementing this feature naively would cause cyclic object transfer dependencies. For example, it is possible that two nodes try to fetch the same object from each other. It is because when the a receiver locates an alternative sender, the object directory can return the address of another node which fetches the object from the receiver. To avoid cyclic dependencies, we need to track the dependencies of \texttt{Get} if the sender is not the original task that creates the object. If a sender fails, the receiver only resumes if it can find another sender whose dependencies do not include the receiver itself. \autorefsuffix{fig:multicast}{c'} shows the previous example if \texttt{R1} fails. \texttt{R2} resumes the fetch from \texttt{S}, and when \texttt{R3} comes, \texttt{R3} can fetch from \texttt{R2} (\autorefsuffix{fig:multicast}{d'}).

\subsubsection{Reduce} When a task fails during \texttt{Reduce}, this node is immediately removed from the tree by the coordinator, and will be replaced by the next ready source object. The guarantee is that to reduce $n$ objects from $m$ source objects, as long as at least $n$ objects can be created (i.e., $m-n$ tasks can fail), \texttt{Reduce} will return successfully. Otherwise, \texttt{Reduce} completes when enough failed tasks are reconstructed by the underlying task-based system's recovery mechanisms. A failed tree node causes its parent, its grandparent, and all its ancestors to clear the reduced object. In the previous example, \autoref{fig:reduce-tree-faulttolerant} shows the adapted tree after \texttt{R2} fails. If the task \texttt{Reduce} 6 out of 10 objects and  \texttt{R2} is recovered after \texttt{R7} arrives, \texttt{R7} replaces \texttt{R2}'s position in the tree. (\texttt{R7} can also be the rejoined \texttt{R2}.) \texttt{R4} has to clear all the current reduced the object, because the final result should be the \texttt{Reduce} result of \texttt{R1}, \texttt{R3}, \texttt{R4}, ..., \texttt{R7}. Any intermediate result that contains \texttt{R2}'s object has to be cleared. 
Overall, at most $\log_d n$ nodes have to clear the current object.

\section{Implementation}
The core of \codename is implemented using 3957 lines of C++. We provide a C++ and a Python front-end. The Python front-end is implemented using 645 lines of Python and 275 lines of Cython. We build the Python front end because it is easier to integrate with Ray~\cite{moritz2018ray} and other data processing libraries (e.g.,  Numpy~\cite{numpy}, TensorFlow~\cite{abadi2016tensorflow}, PyTorch~\cite{paszke2019pytorch}). The interface between the Python front-end and the C++ backend is the same as \codename's API (\autoref{tab:api}).

We implement the object directory service using a set of gRPC~\cite{grpc} server processes distributed across nodes. Each directory server can push location notifications directly to an object store node. Each object store node in \codename is a gRPC server with locally buffered objects. Upon a transfer request from a remote node (e.g., during \texttt{Get}), the node sets up a direct TCP connection to the remote node and pushes the object buffer through the TCP connection.

In our experiments, we observe that setting $d$ to $1, 2,$ or $n$ in the tree reduce algorithm is enough for our applications. When a task calls \texttt{Reduce}, \codename picks $d$ from $1, 2$ and $n$ that minimizes the estimated total latency based on the network latency $L,$ bandwidth $B,$ and the object size $S$. \autoref{sec:ablation} shows the effect of different choices of $d$.

\section{Evaluation}
\label{sec:eval}
We first microbenchmark \codename on a set of popular traditional network primitives (e.g., broadcast, reduce, allreduce). We then evaluate \codename using real applications on Ray~\cite{moritz2018ray}, including asynchronous SGD, reinforcement learning, and serving an ensemble of ML models. We also test \codename with synchronous data-parallel training workloads to estimate how much performance is lost if people choose to run these static and synchronous workloads on task-based distributed systems. Each application requires <100 lines of code changes, most of which are for object serialization. All experiments are done on AWS EC2. We use a cluster of 16 m5.4xlarge nodes (16 vCPUs, 64GB memory, 10\,Gbps network) with Linux (version 4.15). We run every test 10 times, and we show standard deviations as error bars.

\begin{figure}[t]
\centering
\begin{subfigure}[b]{0.15\textwidth}
\centering
\includegraphics[height=2.5cm]{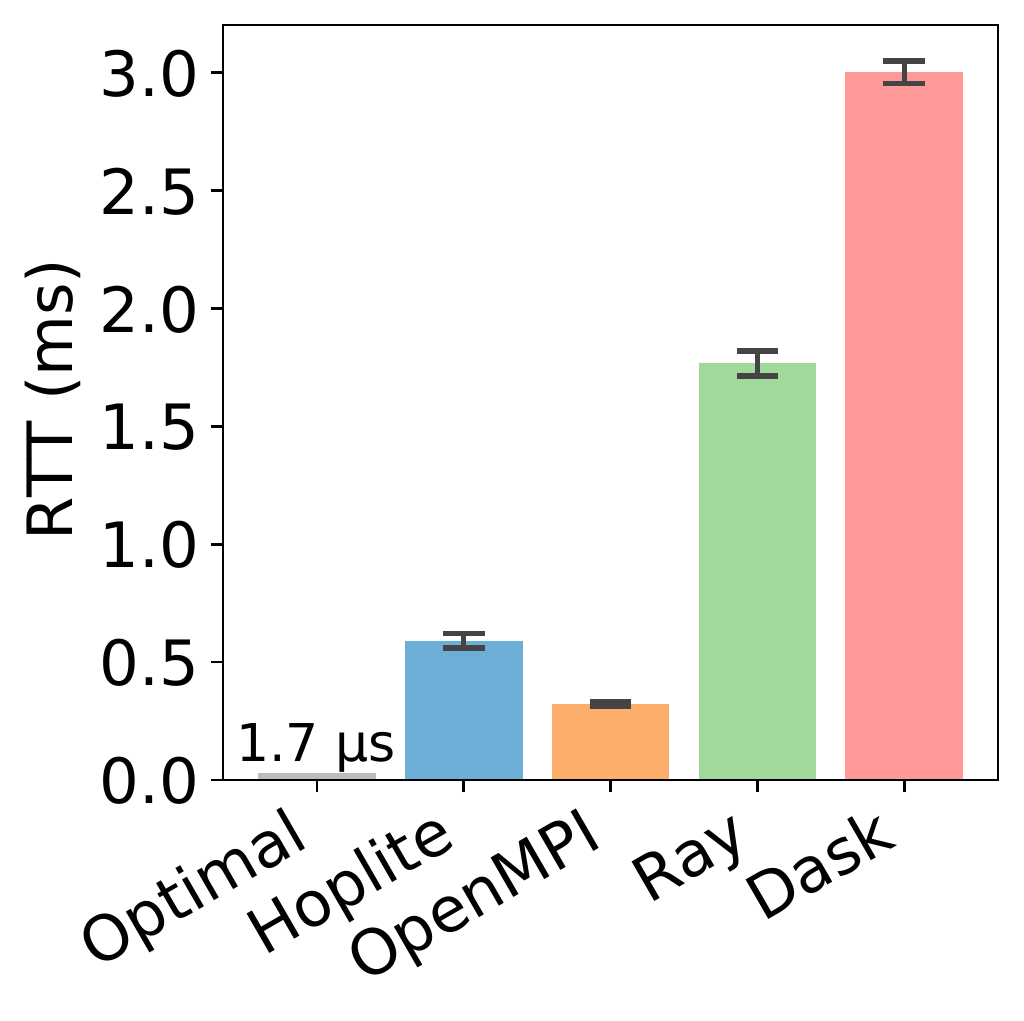}
\label{fig:rtt1k}
\subcaption{1\,KB}
\end{subfigure}
\begin{subfigure}[b]{0.15\textwidth}
\centering
\includegraphics[height=2.5cm]{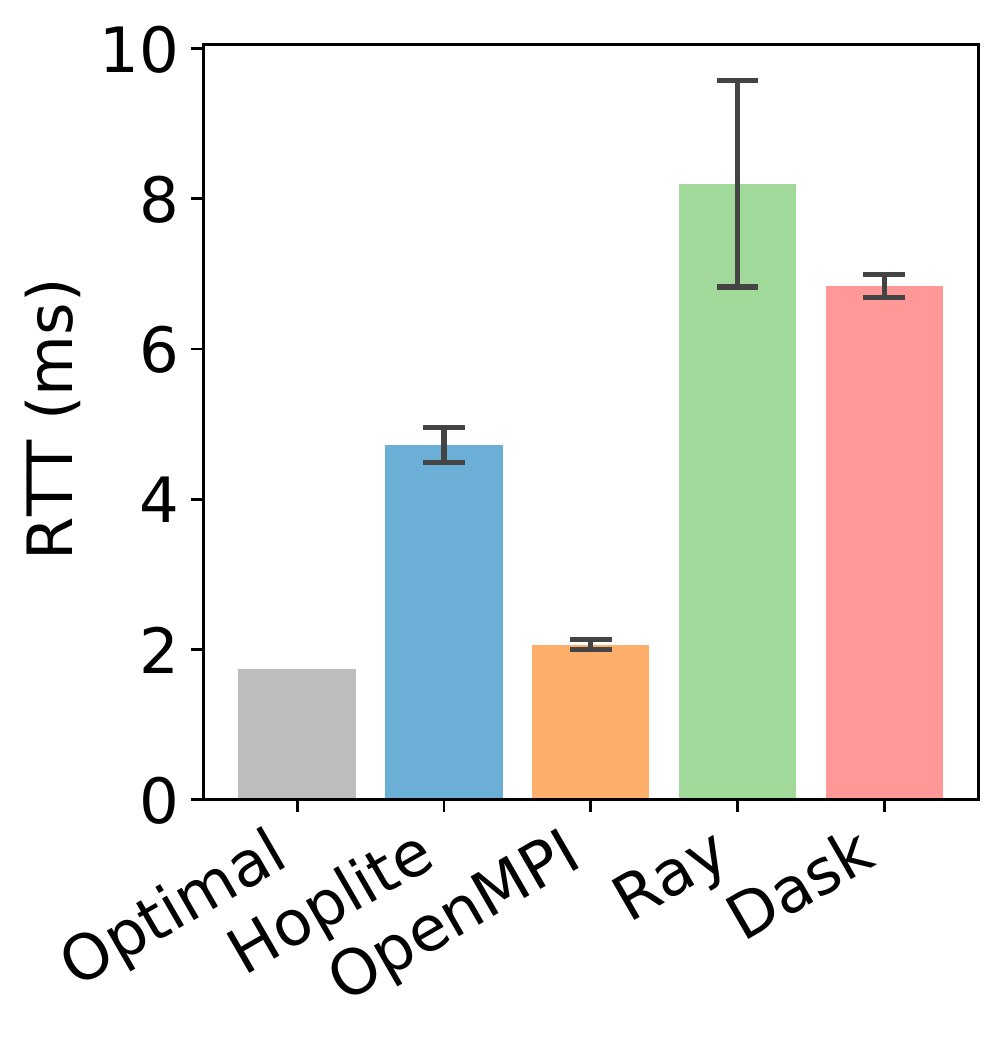}
\label{fig:rtt1m}
\subcaption{1\,MB}
\end{subfigure}
\begin{subfigure}[b]{0.15\textwidth}
\centering
\includegraphics[height=2.5cm]{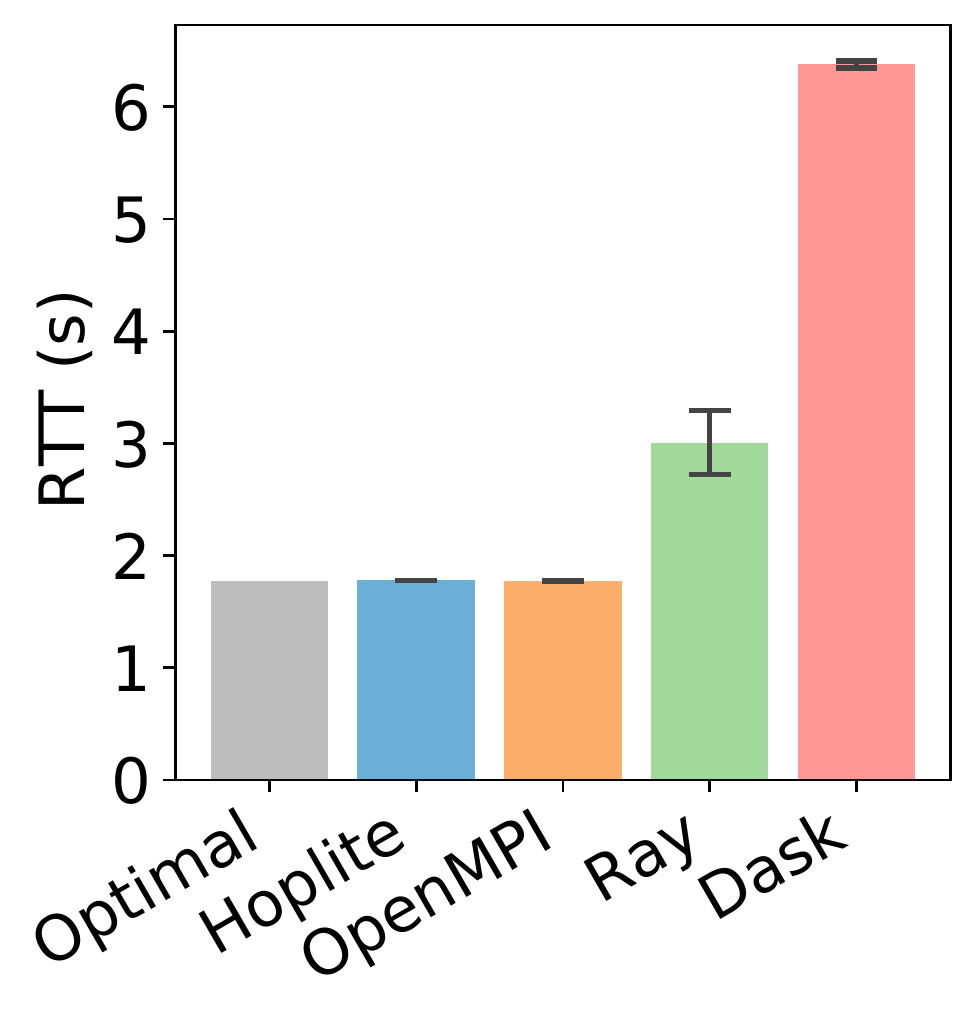}
\label{fig:rtt1g}
\subcaption{1\,GB}
\end{subfigure}
\caption{\normalfont \small Round trip latency for point-to-point data communication on \codename, OpenMPI, Ray, and Dask. We also include the theoretical optimal RTT (i.e. total bytes transferred divided by the bandwidth).}
\label{fig:rtt}
\end{figure}

\subsection{Microbenchmarks}
\label{sec:microbenchmarks}

\begin{figure*}[t]
\centering
\includegraphics[width=\textwidth]{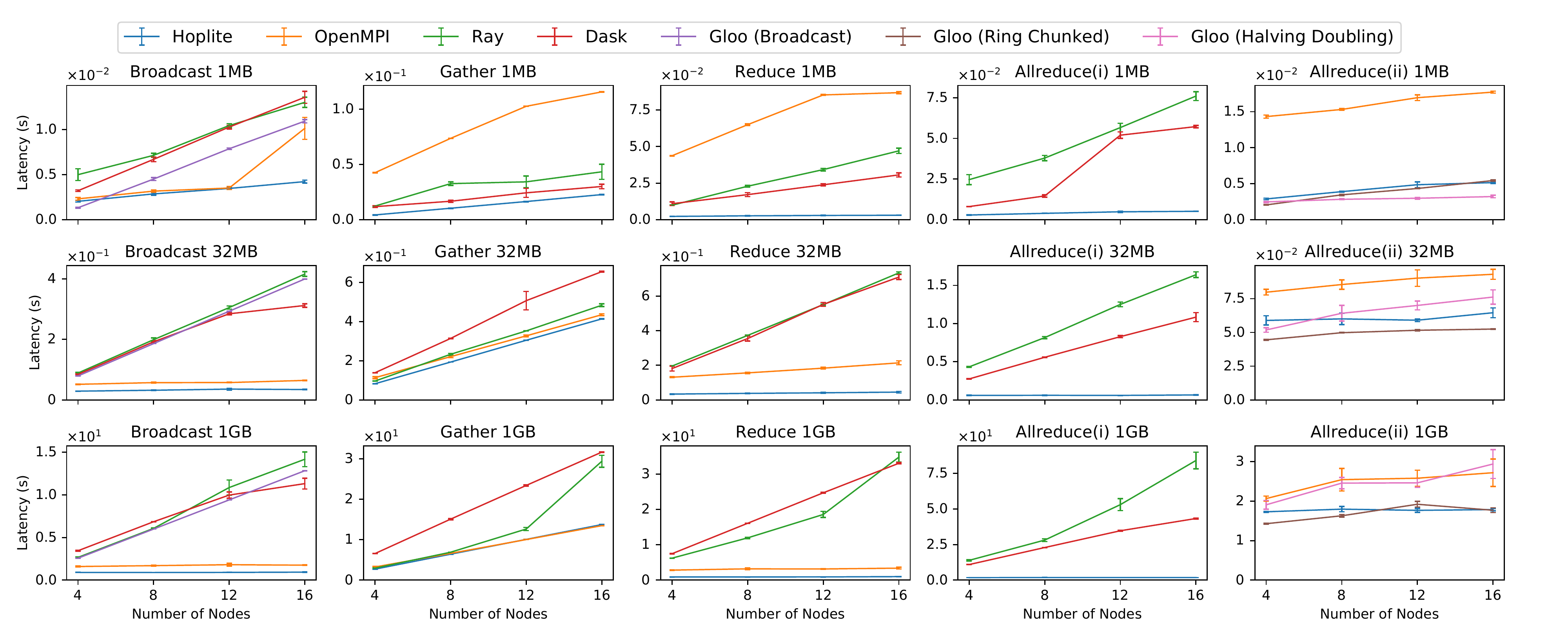}
\caption{\normalfont \small Latency comparison of \codename, OpenMPI, Ray, Dask, and Gloo on standard collective communication primitives (e.g., broadcast, gather, reduce, allreduce). To show the results more clearly, we split the results of Allreduce into two groups: group (i) includes \codename, Ray, and Dask, and group (ii) includes \codename, OpenMPI, and two different allreduce algorithms in Gloo.}
\label{fig:collective_communication}
\end{figure*}

We use two popular task-based distributed systems,  Ray~\cite{moritz2018ray} (version 0.8.6) and Dask~\cite{rocklin2015dask} (version 2.25), as our baselines. In addition, we compare \codename with OpenMPI~\cite{graham2005open} (version 3.3) and Gloo~\cite{gloo}. We chose OpenMPI because OpenMPI is the collective communication library recommended by AWS. We did not choose Horovod because Horovod has three backends: OpenMPI, Gloo, and NCCL. We have already tested OpenMPI and Gloo individually. We currently do not support GPU, so we do not test NCCL.

\subsubsection{Point-to-Point Data Communication}

We first benchmark direct point-to-point transfer. On our testbed, writing object locations to the object directory service takes 167\,\si\micro s (standard deviation = 12\,\si\micro s), and getting object location from the object directory service takes 177\,\si\micro s (standard deviation = 14\,\si\micro s).

\codename's point-to-point communication is efficient. We test round-trip time for different object sizes using OpenMPI, Ray, Dask, and \codename. \autoref{fig:rtt} shows the result. We also include the optimal RTT, which is calculated by ${\text{object\_size}}/{\text{bandwidth}} \times 2$.

For 1\,KB and 1\,MB object, OpenMPI is 1.8x and 2.3x faster than \codename. For 1\,GB objects, \codename is 0.2\% slower than OpenMPI. Ray and Dask are significantly slower. OpenMPI is the fastest because MPI has the knowledge of the locations of the processes to communicate. Ray, Dask, and \codename need to locate the object through an object directory service. \codename outperforms Ray and Dask because (1) \codename stores object contents in object directory service for objects smaller than 64\,KB (\autoref{sec:directoryservice}) and (2) \codename uses pipelining (\autoref{sec:pipelining}) to reduce end-to-end latency. Ray does not support pipelining, so it suffers from the extra memory copy latency in the object store. Our pipelining block size is 4\,MB, and thus larger object (1\,GB) has better pipelining benefits. On 1\,GB object, \codename achieves similar performance as the underlying network bandwidth despite it has additional memory copies. This is because fine-grained pipelining successfully overlaps memory copying and data transfer.

\subsubsection{Collective Communication}

Next, we measure the performance of collective communication on OpenMPI, Ray, Dask, Gloo, and \codename, with arrays of 32-bit floats and addition as the reduce operation (if applicable). We measure the time between when the input objects are ready and when the last process finishes. For both \codename and Ray, we assume that the application uses a read-only \texttt{Get} to avoid the memory copy from the object store to the receiver task (\autoref{sec:pipelining}). Gloo only implements broadcast and allreduce. For allreduce, Gloo supports several algorithms. We evaluated the performance for all of them, and for presentation simplicity, we only show the two algorithms with the best performance on our setup: (1) ring-chunked allreduce and (2) halving doubling allreduce.

\begin{figure*}[t]
    \centering
\begin{subfigure}[b]{0.23\textwidth}
\includegraphics[width=\textwidth]{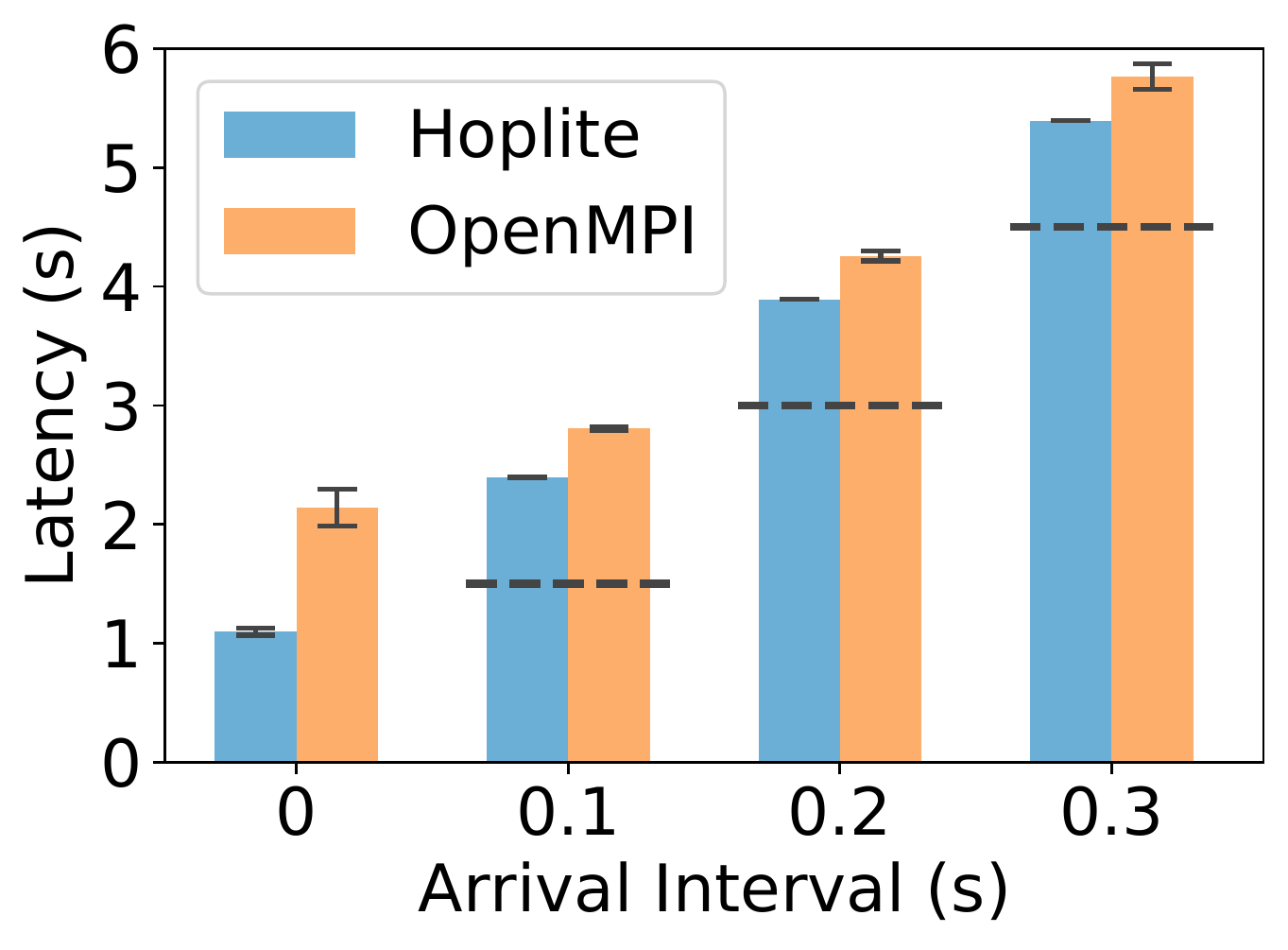}
\label{fig:delay_broacast}
\subcaption{Broadcast}
\end{subfigure}
\hspace{6mm}
\begin{subfigure}[b]{0.23\textwidth}
\includegraphics[width=\textwidth]{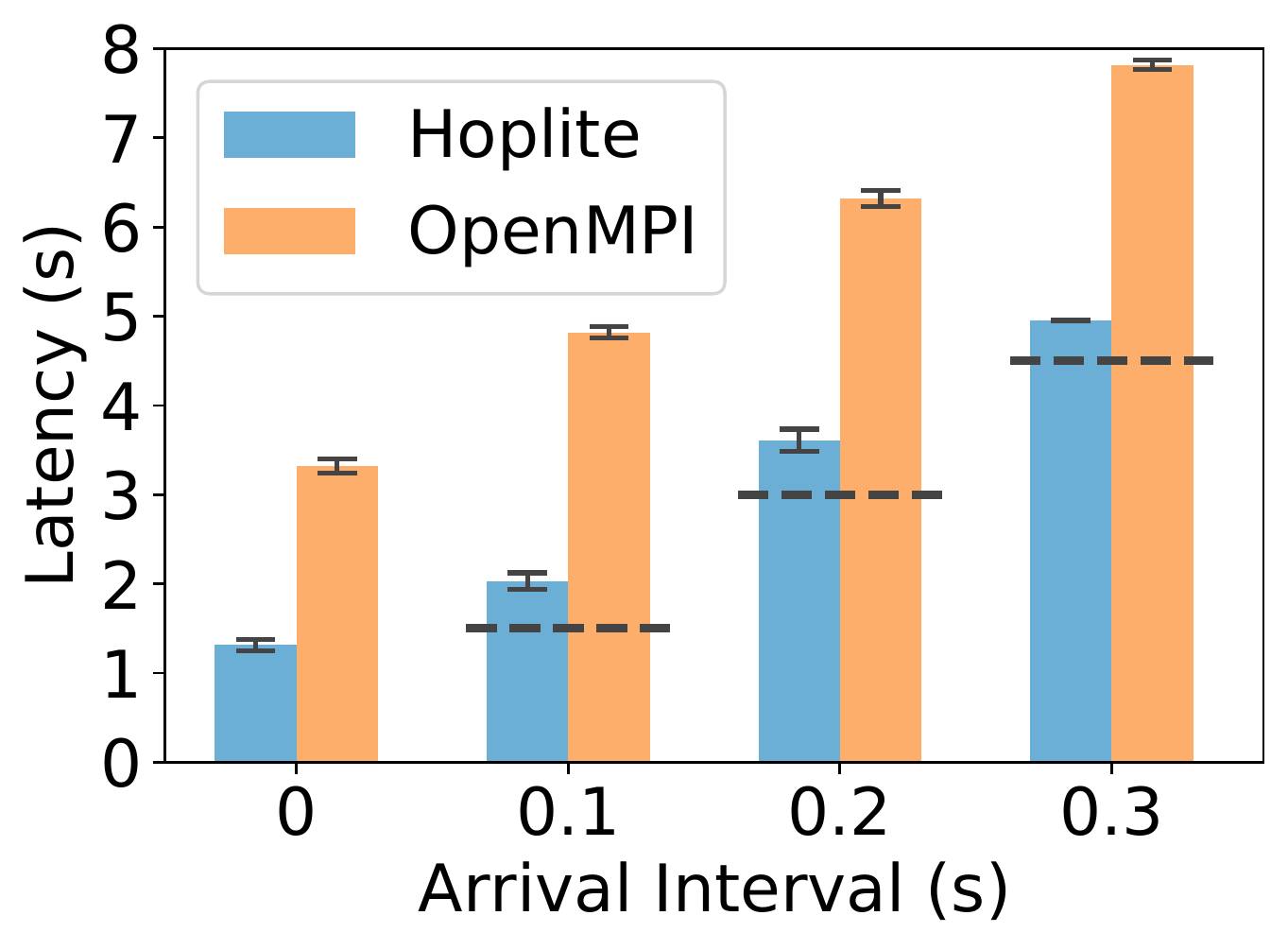}
\label{fig:delay_reduce}
\subcaption{Reduce}
\end{subfigure}
\hspace{6mm}
\begin{subfigure}[b]{0.23\textwidth}
\includegraphics[width=\textwidth]{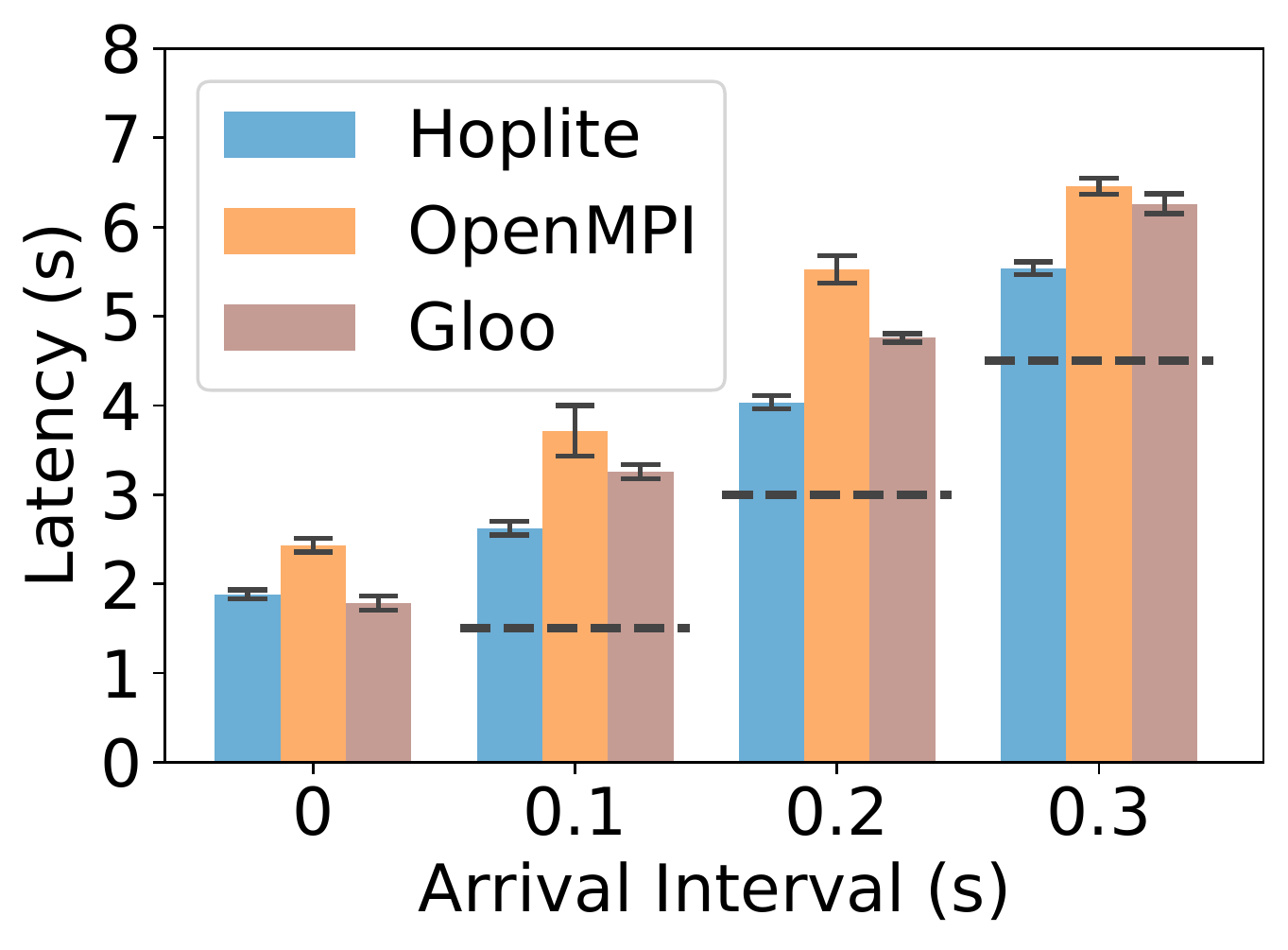}
\label{fig:delay_allreduce}
\subcaption{AllReduce}
\end{subfigure}
    \caption{\normalfont \small Latency of 1\,GB object broadcast/reduce/allreduce on 16 nodes when tasks start sequentially with a fixed arrival interval. Arrival interval equals to 0 means that all the tasks start at the same time. The dashed lines denote the time the last task arrives.}
    \label{fig:delaystart}
\end{figure*}

\autoref{fig:collective_communication} shows the results for medium (1MB) to large (1GB) objects.\footnote{OpenMPI's latency does not increase monotonically because OpenMPI chooses different algorithms on different conditions (e.g., number of nodes, whether the number of nodes is a power of two, object size).} We present the results for small objects (1KB, 64KB) in \autoref{sec:smallobject} because small objects are cached in object directory service in \codename, and there is thus no collective communication to begin with. In summary, \codename achieves a similar level of performance as traditional collective communication libraries, such has OpenMPI and Gloo. \codename significantly outperforms Ray and Dask, 
because Ray and Dask do not support efficient collective communication. Gloo's ring-chunked allreduce is the fastest allreduce implementation for large objects in our tests.

\emph{Broadcast.} We let one node first \texttt{Put} an object, and after the \texttt{Put} succeeds, other nodes \texttt{Get} the object simultaneously. The latency of broadcast is calculated from the time all nodes call \texttt{Get} to the time when the last receiver finishes. \codename and OpenMPI achieve the best performance for all object size and node configurations. This is because Ray, Dask, and Gloo do not have collective communication optimization for broadcast. \codename slightly outperforms OpenMPI because of fine-grained pipelining.

\emph{Gather.} We let every node first \texttt{Put} an object, and after every node's \texttt{Put} succeeds, one of the nodes \texttt{Get} all the object via their \texttt{ObjectID}s. The latency of gather is the \texttt{Get} duration. OpenMPI and \codename outperforms the rest for all object size and node configurations. This is because both Ray and Dask need additional memory copying between workers and the object store. \codename also needs additional memory copying, but the latency is masked by fine-grained pipelining between workers and the object store.

\emph{Reduce.} We let every node first \texttt{Put} an object, and after every node's \texttt{Put} succeeds, one of the nodes \texttt{Reduce} the objects via their \texttt{ObjectID}s to create a new \texttt{ObjectID} for the result. The node then calls \texttt{Get} to get the resulting object buffer. The latency of reduce is calculated from the time the node calls \texttt{Reduce} to the time the node has a copy of the reduce result. OpenMPI and \codename consistently outperform the rest for all object size and node configurations since Ray and Dask do not support collective communication. \codename can slightly outperform OpenMPI because of fine-grained pipelining.

\emph{AllReduce.} In \codename, we simply concatenate reduce and broadcast to implement allreduce. The latency of allreduce is calculated from the time a node starts to \texttt{Reduce} all the objects to the last node \texttt{Get} the reduce result. We divide the results into two groups in \autoref{fig:collective_communication}. \codename significantly outperforms Ray and Dask because of the collective communication support of broadcast and reduce in \codename. Note that efficient allreduce is not our design goal since allreduce is a static and synchronous collective communication operation. However, \codename still achieves comparable performance with static collective communication libraries such as OpenMPI and Gloo.

\subsubsection{Asynchrony}
\codename's performance is robust even when processes are not synchronized, which is typical in task-based distributed systems. We measure broadcast, reduce, and allreduce latencies when the participating tasks arrive sequentially with a fixed arrival interval. For broadcast (\autorefsuffix{fig:delaystart}{a}), OpenMPI makes some progress before the last receiver arrives (\autoref{sec:related}). However, the algorithm is static (i.e. based on process \emph{rank} \cite{graham2005open}), while \codename achieves a lower latency with a dynamic algorithm that does not depend on the particular arrival order. We do not include Gloo because it does not optimize its broadcast performance (\autoref{fig:collective_communication}). For reduce (\autorefsuffix{fig:delaystart}{b}) and allreduce (\autorefsuffix{fig:delaystart}{c}), both OpenMPI and Gloo have to wait until all processes are ready, while \codename can make significant progress before the last object is ready. This allows \codename to even outperform Gloo's ring-chunked allreduce when objects do not arrive at the same time.

\begin{figure}[t]
    \centering
\begin{subfigure}[b]{0.23\textwidth}
\includegraphics[width=\textwidth]{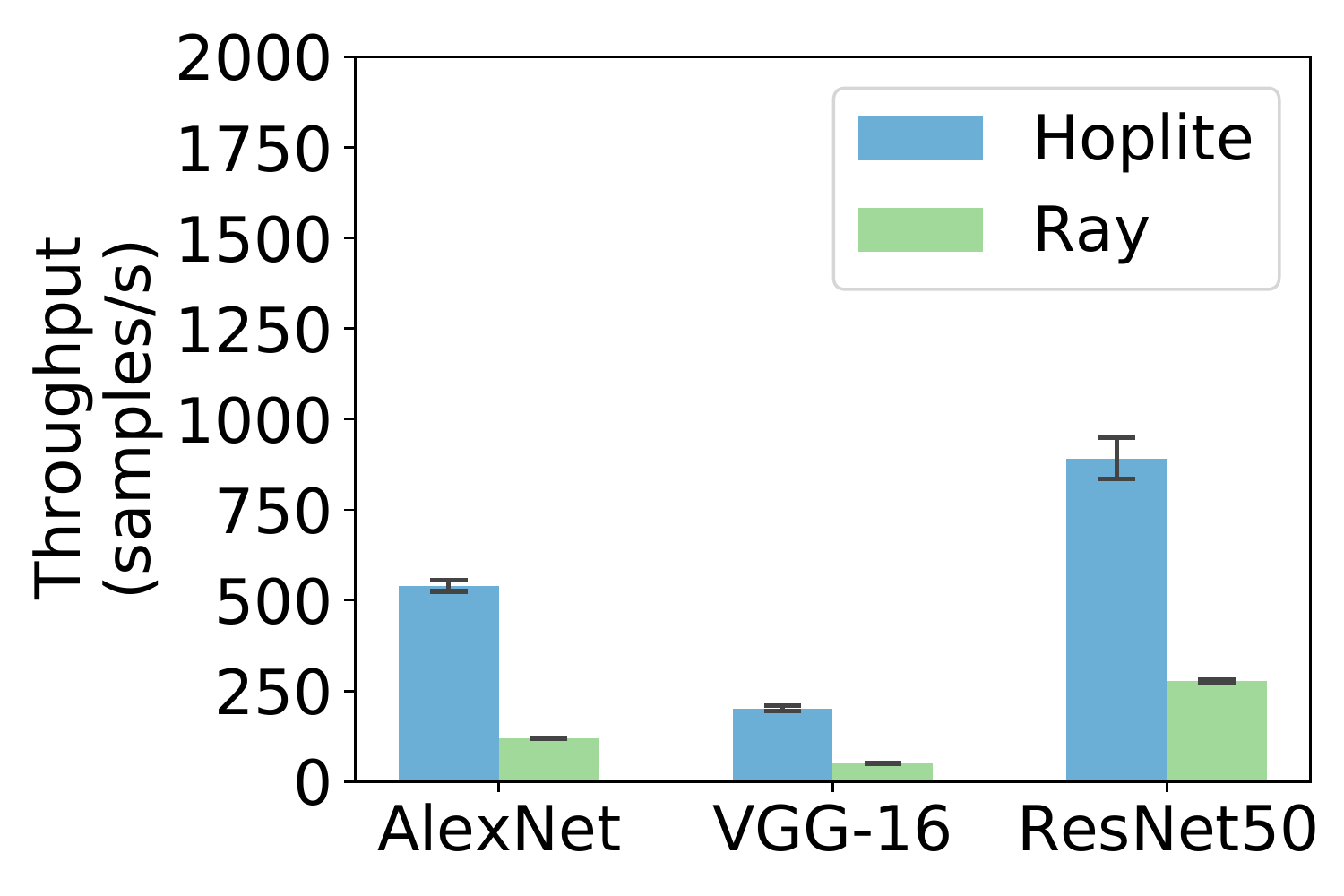}
\subcaption{Number of Nodes = 8}
\end{subfigure}
\begin{subfigure}[b]{0.23\textwidth}
\includegraphics[width=\textwidth]{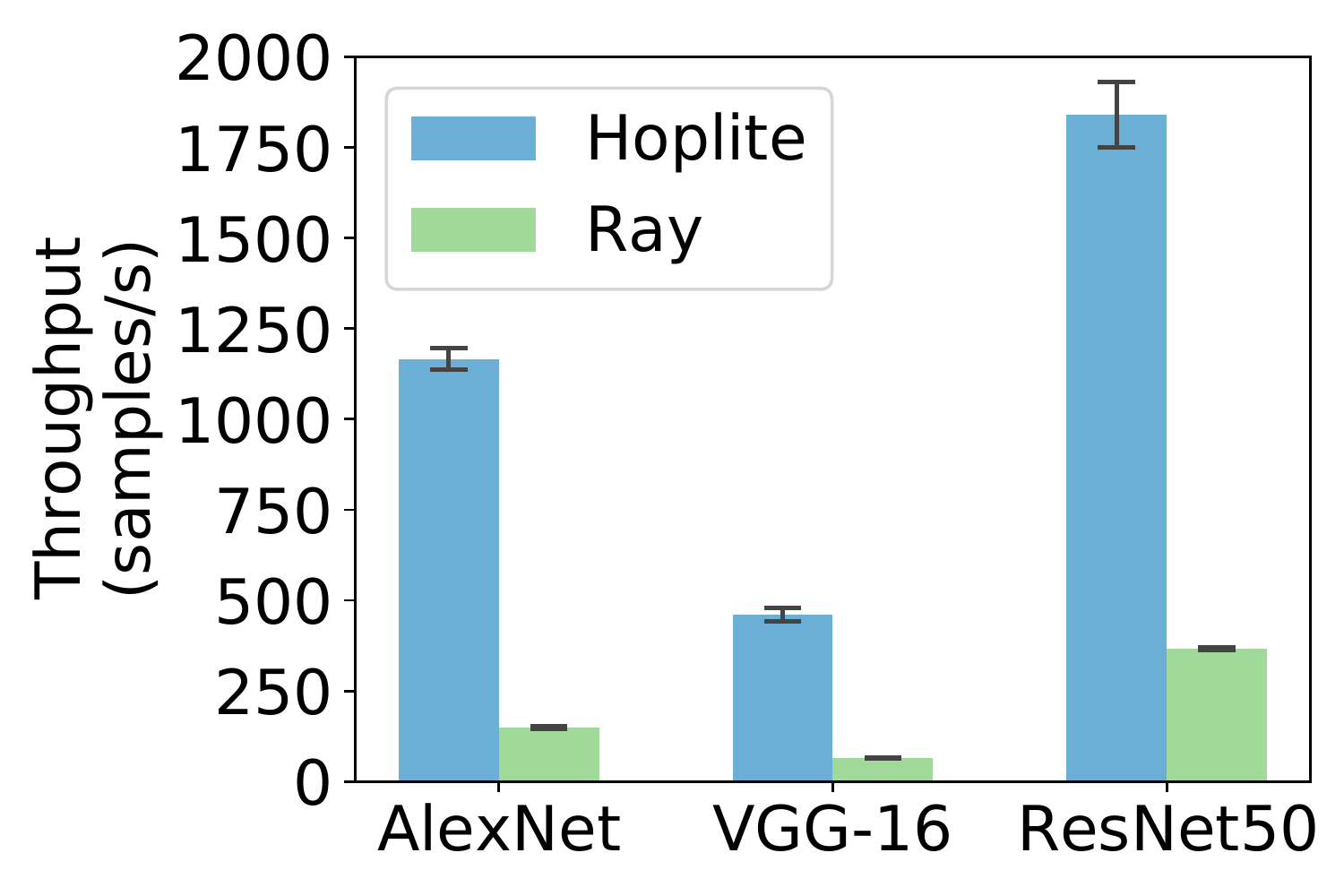}
\subcaption{Number of Nodes = 16}
\end{subfigure}
\caption{\normalfont \small Training
throughput (number of training samples per second) for asynchronous SGD.}
\label{fig:asyncps}
\end{figure}

\subsection{Asynchronous SGD}
Asynchronous stochastic gradient descent (SGD) is one way to train deep neural networks efficiently, and it usually uses a parameter server framework \cite{li2013parameter, li2014scaling, dean2012large, li2013parameter, li2014scaling}: clients fetch the parameters from a centralized server, evaluate the parameters on its own portion of data (e.g., performing forward and backward propagation on a neural network), and send the updates (e.g., gradients) back to the server independently. The parameter server needs to broadcast parameters to and reduce from an uncertain set of workers. 

Here we evaluate \codename with Ray's example implementation of an asynchronous parameter server~\cite{rayps}. We use three widely-used standard deep neural networks, AlexNet~\cite{krizhevsky2012imagenet} (model size = 233\,MB), VGG-16~\cite{simonyan2014very} (model size = 528\,MB), and ResNet-50~\cite{he2016deep} (model size = 97\,MB). We test two cluster configurations: 8 p3.2xlarge nodes and 16 p3.2xlarge nodes on AWS. p3.2xlarge instance has the same network performance as m5.4xlarge instance but with an additional NVIDIA V100 GPU to accelerate the execution of the neural networks. The asynchronous parameter server collects and reduces the updates from the first half of worker nodes that finish the update and broadcast the new weights back to these nodes.

We show the results in \autoref{fig:asyncps}. \codename improves the training throughput of the asynchronous parameter server. Comparing to Ray, it speedups training on asynchronous parameter server for 16 nodes by 7.8x, 7.0x, and 5.0x, for AlexNet, VGG-16, and ResNet-50, respectively. Ray is slow because the parameter server has to receive gradients from each worker and send the updated model to each worker one by one. This creates a bandwidth bottleneck at the parameter server. In \codename, these operations are optimized by our broadcast and reduce algorithms.

\subsection{Reinforcement Learning}
RL algorithms involve the deep nesting of irregular distributed computation patterns, so task-based distributed systems are a perfect fit for these algorithms. We evaluate \codename with RLlib~\cite{liang2018rllib}, a popular and comprehensive RL library on Ray. Distributed RL algorithms can be divided into two classes: In \emph{samples optimization} (e.g., IMPALA~\cite{espeholt2018impala}, Asynchronous PPO~\cite{schulman2017proximal}), a centralized trainer periodically broadcasts a policy to a set of workers and gather the rollouts generated by the workers to update the model. In \emph{gradients optimization} (e.g., A3C~\cite{Mnih2016a3c}), the workers compute the gradient with their rollouts, and the trainer updates the model with the reduced gradients from the workers.

\begin{figure}[t]
\centering
\begin{subfigure}[b]{0.23\textwidth}
\includegraphics[width=\textwidth]{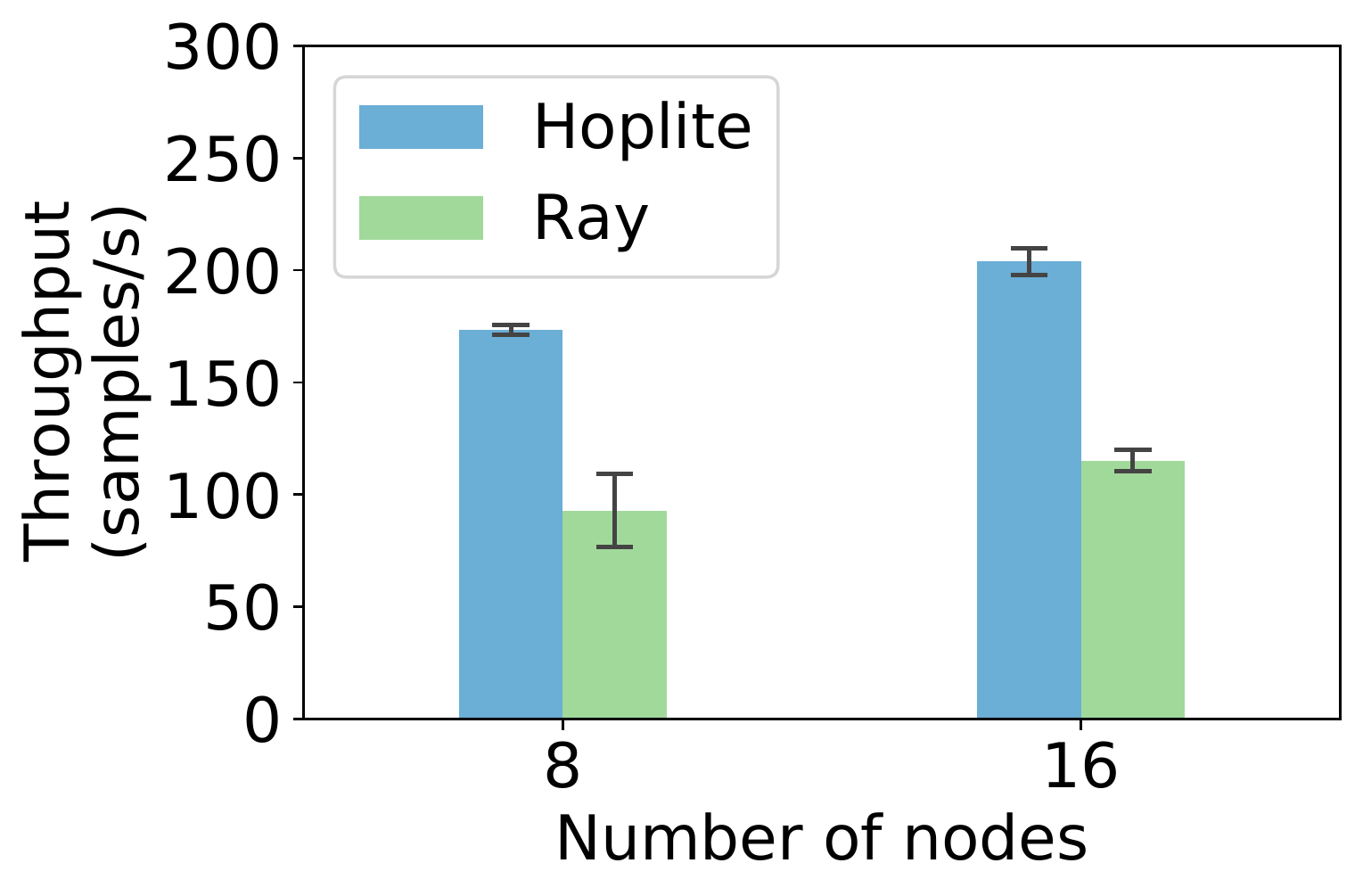}
\subcaption{IMPALA}
\label{fig:impala}
\end{subfigure}
\begin{subfigure}[b]{0.23\textwidth}
\includegraphics[width=\textwidth]{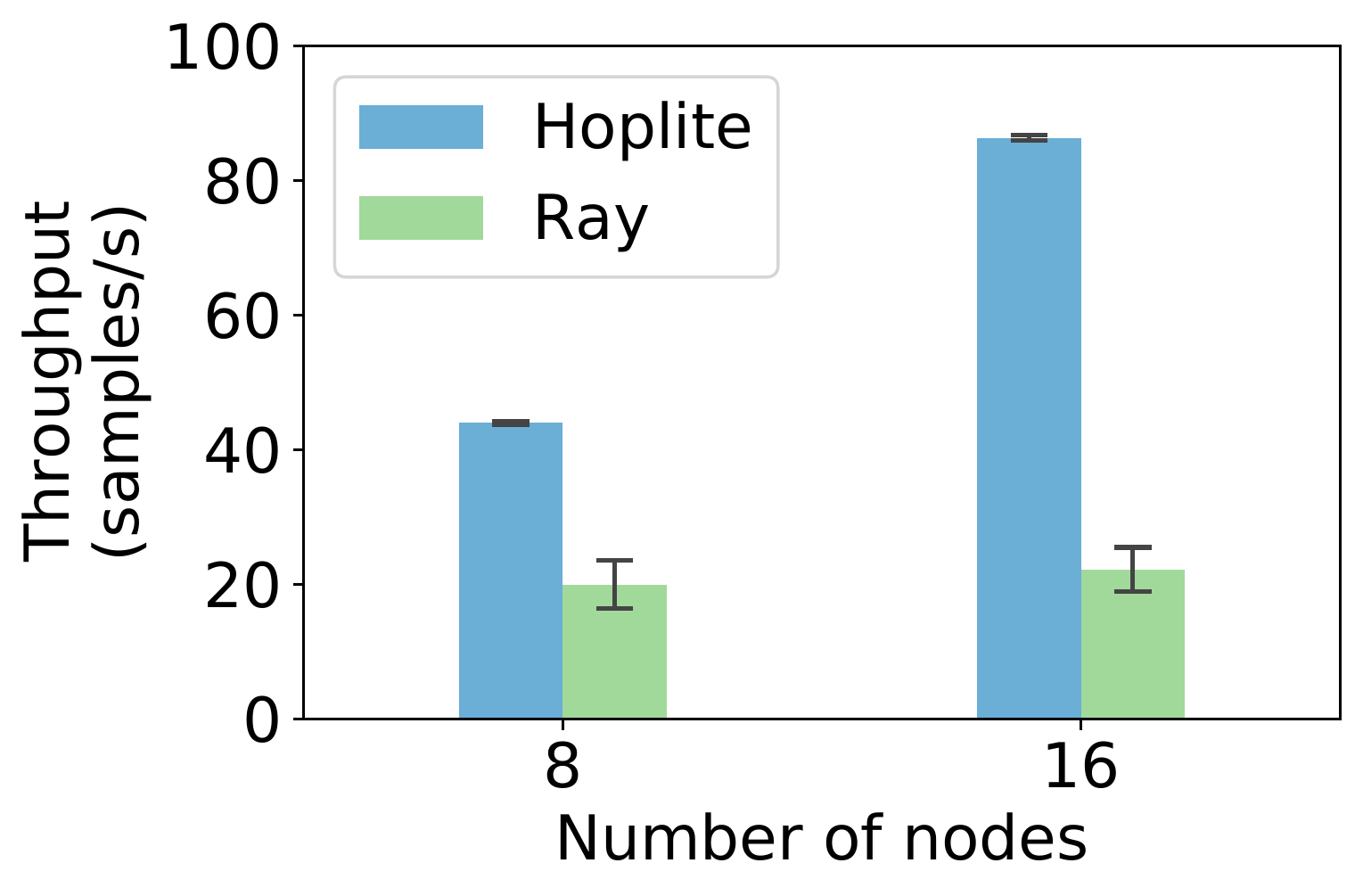}
\subcaption{A3C}
\label{fig:a3c}
\end{subfigure}
\caption{\normalfont \small RLlib's training throughput (number of training samples per second) on Ray and \codename.} 
\label{fig:rllib}
\end{figure}

We evaluate two popular RL algorithms, IMPALA~\cite{espeholt2018impala} and A3C~\cite{Mnih2016a3c}, one from each class. We test two cluster configurations: 8 nodes (1 trainer + 7 workers) and 16 nodes (1 trainer + 15 workers). The trainer broadcast a model to the first half workers that have finished a round of simulation (in IMPALA) or gradient computation (in A3C). We use a two-layer feed-forward neural network with 64\,MB of parameters. \autoref{fig:rllib} shows the training throughput. Training throughput is calculated by the number of simulation traces (in samples optimization) or gradients (in gradients optimization) the RL algorithm can process in a second.

\codename significantly improves the training throughput of both IMPALA and A3C. \codename improves the training throughput of IMPALA by 1.9x on an 8-node cluster and 1.8x on a 16-node cluster. The reason \codename outperforms Ray is because IMPALA has to broadcast a model of 64\,MB frequently to the workers. We expect more improvement when the number of nodes is higher, but we already achieve the maximum training throughput: IMPALA is bottlenecked by computation rather than communication using \codename with 16 nodes (15 workers). For A3C, \codename improves the training throughput by 2.2x on the 8-node configuration and 3.9x on the 16-node configuration. Unlike IMPALA, A3C achieves almost linear scaling with the number of workers. A3C on Ray cannot scale linearly from 8 nodes to 16 nodes because of the communication bottleneck.

\subsection{ML Model Serving}
\label{sec:serve}
Machine learning is deployed in a growing number of applications which demand real-time, accurate, and robust predictions under heavy query load \cite{crankshaw2017clipper, baylor2017tfx, olston2017tensorflow}. An important use case of task-based distributed system is to serve a wide range of machine learning models implemented with different machine learning frameworks \cite{moritz2018ray}. 

We evaluate \codename with Ray Serve \cite{rayserve}, a framework-agnostic distributed machine learning model serving library built on Ray. We set up an image classification service with a majority vote-based ensemble of the following models: AlexNet \cite{krizhevsky2012imagenet}, ResNet34 \cite{he2016deep}, EfficientNet-B1/-B2 \cite{tan2019efficientnet}, MobileNet V2 \cite{sandler2018mobilenetv2}, ShuffleNet V2 x0.5/x1.0 \cite{ma2018shufflenet}, and SqueezeNet V1.1 \cite{iandola2016squeezenet}. We test two cluster configurations: 8 p3.2xlarge nodes and 16 p3.2xlarge nodes on AWS. For 8 nodes setting, we serve a different model on each node. For 16 nodes setting, each model is served by two different nodes and the two nodes serve the model with two different versions of weight parameters. Each query to the service includes a batch of 64 images of size 256$\times$256. During serving, the service will broadcast the query to all the nodes to evaluate on different models, gather the classification results, and return the majority vote to the user.

\begin{figure}[t]
\centering
\includegraphics[width=0.25\textwidth]{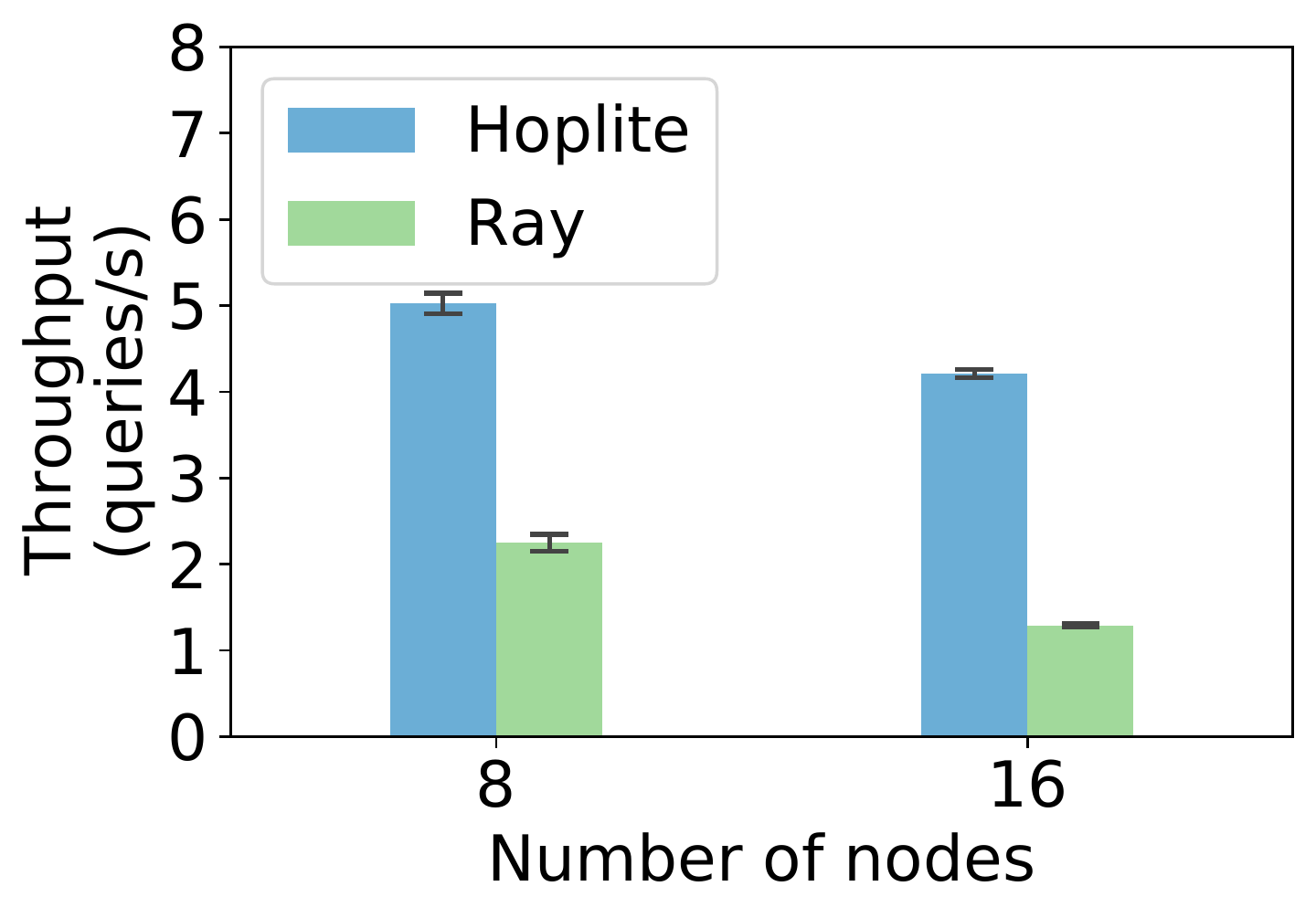}
\caption{\normalfont \small Ray Serve's performance (queries per second) on Ray and \codename for an ensemble of image classification models.} 
\label{fig:serve}
\end{figure}

We visualize the results in \autoref{fig:serve}. \codename improves the serving throughput for serving an ensemble of image classification models. Comparing to Ray, it speedups the serving throughput by 2.2x for 8 nodes and 3.3x for 16 nodes. This shows that the optimized broadcast algorithm in \codename helps Ray Serve to improve the serving throughput.

\begin{figure}[t]
\centering
\begin{subfigure}[b]{0.23\textwidth}
\includegraphics[width=\textwidth]{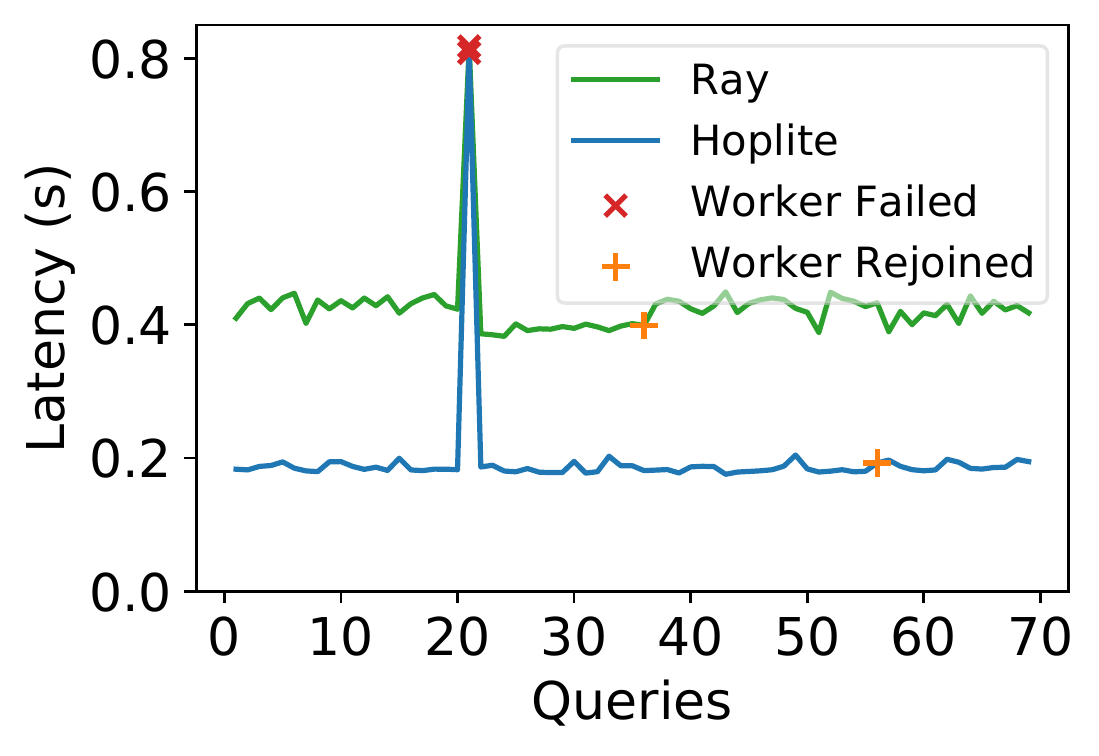}
\subcaption{Ray Server latency.}
\label{fig:ray_serve_fault}
\end{subfigure}
\begin{subfigure}[b]{0.23\textwidth}
\includegraphics[width=\textwidth]{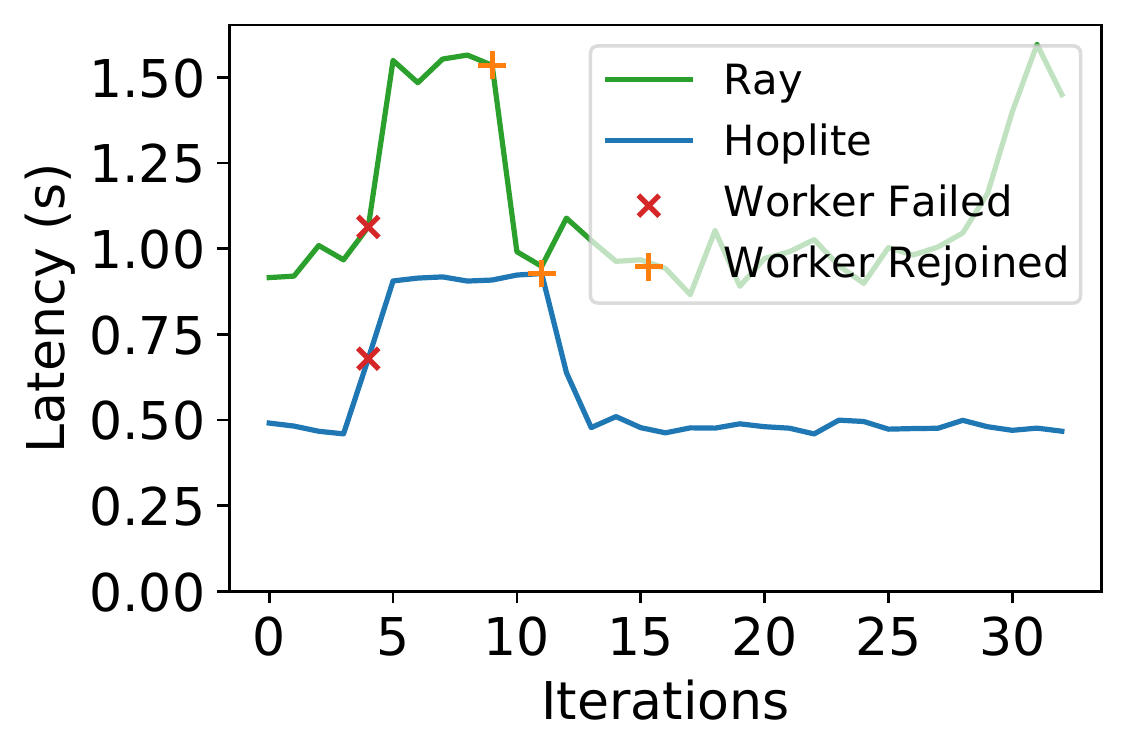}
\subcaption{Async SGD latency.}
\label{fig:aysync_sgd_fault}
\end{subfigure}
\caption{\normalfont \small Latency when a pariticipating task fails and rejoins on (a) Ray Serve and (b) async SGD.} 
\label{fig:faulttolerance}
\end{figure}

\subsection{Fault Tolerance}

We evaluate the failure recovery latency before and after we apply \codename to Ray. We rerun our model serving with 8 models and async SGD workloads with 6 workers, and we manually trigger a failure. We do this experiment 10 times. \autoref{fig:faulttolerance} shows one particular run. The y-axis shows the latency per query (in model serving) or per iteration (in async SGD), and the x-axis shows the index of the query or the iteration. \codename significantly improves Ray's performance. Ray's failure detection latency is $0.58 \pm 0.13$ second, and after we apply \codename to Ray, Ray's failure detection latency increases to $0.74 \pm 0.05$ second. The additional $28\%$ latency introduced by \codename is because \codename has a different failure detection mechanism. Ray detects failure by monitoring the liveness of the worker process. \codename detects failure by checking the liveness of a socket connection. 

After the failure, Ray Serve's latency drops because it only needs to broadcast to less receivers. The latency comes back to normal after the failed worker rejoins. For \codename, the latency difference is negligible because of the efficient broadcast algorithm. \codename takes more queries between the task fails and the task rejoins. This is because \codename is efficient and has processed more queries during the recovery window (the time between the failure and task rejoin). In async SGD, latency for training each iteration increases in the recovery window because of the temporary loss of a worker. The difference in recovery latency (duration of the recovery window) between Ray and \codename is negligible because both use Ray's mechanism to reconstruct the failed task.

\begin{figure}[t]
    \centering
\begin{subfigure}[b]{0.23\textwidth}
\includegraphics[width=\textwidth]{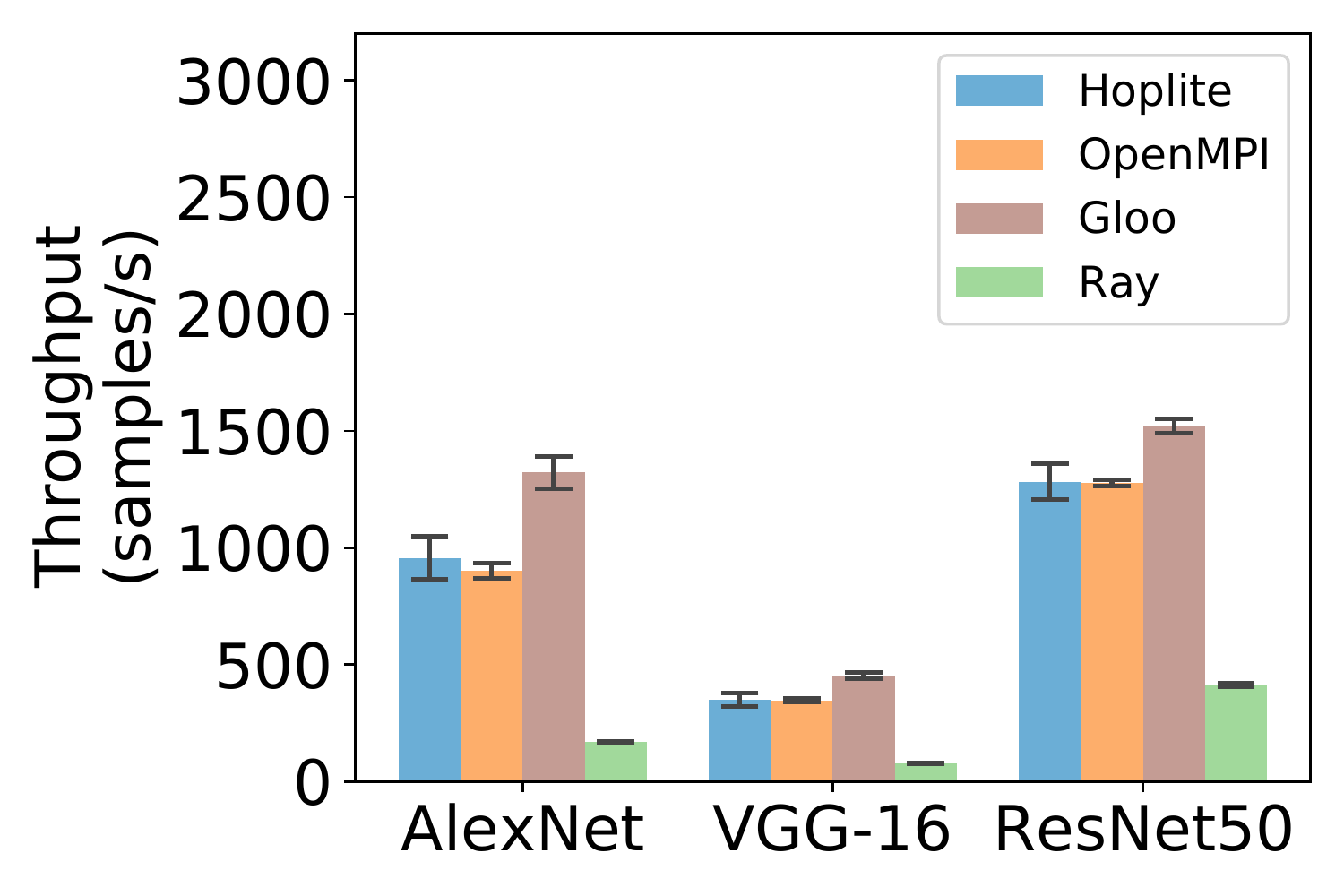}
\vspace{-4mm}
\subcaption{Number of Nodes = 8}
\vspace{-2mm}
\end{subfigure}
\begin{subfigure}[b]{0.23\textwidth}
\includegraphics[width=\textwidth]{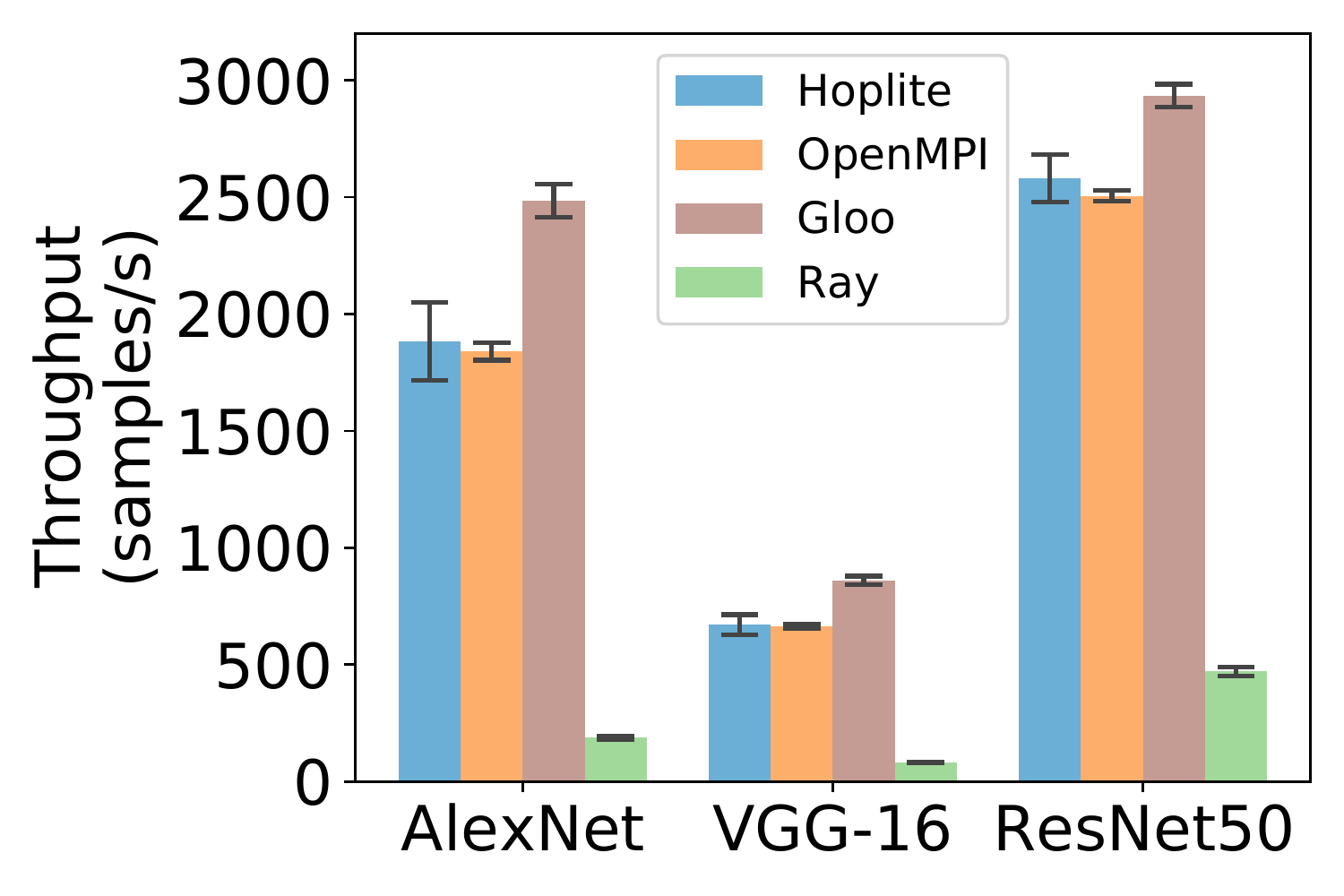}
\vspace{-4mm}
\subcaption{Number of Nodes = 16}
\vspace{-2mm}
\end{subfigure}
    \caption{\normalfont \small Training
    throughput (number of training samples per second) for synchronous data-parallel training.}
\label{fig:syncps}
\end{figure}
\subsection{Synchronous Data-Parallel Training}

Synchronous data-parallel training involves a set of workers, each runs on a partition of training data, and the workers synchronize the gradients each round using allreduce~\cite{goyal2017accurate}. Speeding up synchronous data-parallel training workloads is not our design goal, and they do not require the flexibility provided by task-based systems. Instead, they can run directly on specialized distributed systems that are optimized for static and synchronous workloads (e.g., TensorFlow~\cite{abadi2016tensorflow}, PyTorch~\cite{paszke2019pytorch}). These systems rely on efficient allreduce implementations in traditional collective communication frameworks (e.g., OpenMPI, Gloo). 

However, an interesting question to ask is how much performance developers have to pay if they choose to run these static and synchronous workloads on task-based distributed systems. Our cluster setup is the same as the asynchronous parameter server experiment. In addition to Ray, we evaluate Gloo and OpenMPI. We evaluate the Gloo baseline through PyTorch, which chooses ring-chunked allreduce as its choice for Gloo's algorithm.

We show the results in \autoref{fig:syncps}. \codename significantly improves the synchronous data-parallel training for Ray. Ray is slower than \codename, OpenMPI, and Gloo, with the similar reason as in asynchronous parameter server. \codename achieves similar speed with OpenMPI. However, \codename is 12-24\% slower than Gloo. This is expected because ring-allreduce is more bandwidth efficient than the tree-reduce plus broadcast in \codename.

\section{Discussion}

\emph{Garbage collection.}
\codename provides a \texttt{Delete} call (\autoref{tab:api}) that deletes all copies of an object from the store. This can be used to garbage-collect an object whose \texttt{ObjectID} is no longer in scope in the application. However, it is still the task framework or application's responsibility to determine when \texttt{Delete} can and should be called, since only these layers have visibility into which \texttt{ObjectIDs} a task has references to. The guarantee that \codename provides is simple: when \texttt{Put} is called on an \texttt{ObjectID}, the object copy that is created will be pinned in its local store until the framework calls \texttt{Delete} on the same ID. This guarantees that there will always be at least one available location of the object to copy from, to fulfill future \texttt{Get} requests. Meanwhile, \codename is free to evict any additional copies that were generated on other nodes during execution, to make room for new objects. The overhead of eviction is very low, since \codename uses a local LRU policy per node that considers all unpinned object copies in the local store.

\emph{Framework's Fault tolerance.}
\codename ensures that collective communication can tolerate task failure. A task-based distributed system has a set of control processes that can also fail, and they usually require separate mechanisms to tolerate failures. For example, the object directory service can fail and requires replication for durability. These failures are handled by the underlying framework independent of whether \codename is used.

\emph{Network Heterogeneity.}
The design of \codename assumes that the network capacity between all the nodes is uniform. Accommodating heterogeneous network can achieve higher performance (e.g., using high bandwidth nodes as intermediate nodes for broadcast, fetching objects from a node which has lower latency). This can be done by monitoring network metrics at run time. We do not need this feature for our use cases because our cloud provider ensures uniform network bandwidth between our nodes.

\emph{Integration with GPU.} 
\codename currently does not support pipelining into GPU memory. If training processes need to use GPU, the application has to copy data between GPU and CPU memory. In the future, we want to extend our pipelining mechanism into GPU memory.

\section{Related Work}

\label{sec:related}
\emph{Optimizing data transfer for cluster computing.}
Cluster computing frameworks, such as Spark~\cite{zaharia2016spark} and MapReduce~\cite{dean08mapreduce}, have been popular for decades for data processing, and optimizing data transfer for them~\cite{chowdhury11orchestra, Chowdhury14varys, chowdhury15efficientcoflow, pu15wananalytics, lee2014cloudmirror} has been studied extensively. AI applications are particularly relevant because they are communication-intensive, and traditional collective communication techniques are widely-used~\cite{wang2019blink, alex2018horovod, gloo}. Pipelining is also a well-known technique to improve performance~\cite{narayanan19pipedream, peng2019bytescheduler}. Our work focuses on improving task-based distributed systems~\cite{moritz2018ray, rocklin2015dask, hydro}. Applications on these frameworks have dynamic and asynchronous traffic patterns. To the best of our knowledge, \codename is the first work to provide efficient collective communication support for task-based distributed systems.

\emph{Using named objects or object futures for data communication.}
Using named objects or object futures for data communication is not new. In serverless computing, tasks (or functions) cannot communicate directly. As a result, tasks communicate through external data stores~\cite{pu19locus}, such as Amazon S3~\cite{amazons3} or Redis~\cite{redis}. There, the storage and compute servers are disaggregated, and computer servers do not directly communicate. We target a standard cluster computing scenario, where data is directly transmitted between compute servers. Object futures are a useful construct for expressing asynchronous computation. Dask, Ray, Hydro, and PyTorch~\cite{paszke2019pytorch} all use futures to represent results of remote tasks. Our work is complementary to them, showing that efficient collective communication can co-exist with named objects or object futures.

\emph{Asynchronous MPI.}
MPI supports two flavors of asynchrony. First, similar to a non-blocking POSIX socket, MPI allows an application to issue asynchronous network primitives and exposes an \texttt{MPI\_Wait} primitive to fetch the result. Second, depending on the MPI implementation, some collective communication primitives can make some progress with a subset of participants. For example, in \texttt{MPI\_Bcast}, the sender generates a static broadcast tree. If the receivers arrive in order from the root of the tree to the leaves of the tree, the receivers can make significant progress before the last receiver arrives. If not, then a receiver must wait until all its upstream ancestors are ready before making any progress~(evaluated in \autoref{fig:delaystart}). In \codename, the broadcast tree is generated dynamically at runtime, so the arrival order does not matter. In addition, asynchronous MPI primitives still require applications to specify all the participants before runtime. In \codename, the communication pattern can be expressed dynamically and incrementally, allowing \codename to work with existing task-based distributed systems.

\emph{Collective communication in other domains.}
Optimizing data transfer has been studied extensively in other domains. Application-level multicast~\cite{castro2002multicast, castro2003splitstream} for streaming video on wide-area networks. IP multicast~\cite{ipmulticast} enables a sender to send simultaneously to multiple IP addresses at the same time. These work mostly focus entirely on multicast rather than general-purpose collective communication in distributed computing frameworks.

\section{Conclusion}
Task-based distributed computing frameworks have become popular for distributed applications that contain dynamic and asynchronous workloads. We cannot directly use traditional collective communication libraries in task-based distributed systems, because (1) they require static communication patterns and (2) they are not fault-tolerant. We design and implement \codename, an efficient and fault-tolerant communication layer for task-based distributed systems that achieves efficient collective communication. \codename computes data transfer schedules on the fly, and even when tasks fail, \codename can allow well-behaving tasks to keep making progress while waiting for the failed tasks to recover. We port a popular distributed computing framework, Ray, on top of \codename. \codename speeds up asynchronous SGD, RL, model serving workloads by up to 7.8x, 3.9x, and 3.3x, respectively. \codename's source code is publicly available (\url{https://github.com/suquark/hoplite}). This work does not raise any ethical issues.

\section*{Acknowledgements}

We thank our shepherd Kai Chen and the anonymous reviewers for their insightful feedback. We also thank Hong Zhang and many others at the UC Berkeley RISELab for their helpful discussion and comments. In addition to NSF CISE Expeditions Award CCF-1730628, this research is supported by gifts from Alibaba Group, Amazon Web Services, Ant Group, CapitalOne, Ericsson, Facebook, Futurewei, Google, Intel, Microsoft, Nvidia, Scotiabank, Splunk, and VMware. Danyang Zhuo is supported by an IBM Academic Award.

\bibliographystyle{ACM-Reference-Format}
\bibliography{reference}

\clearpage
\appendix

\section*{Appendix}
Appendices are supporting material that has not been peer-reviewed.

\section{Microbenchmarks on Small Objects}
\label{sec:smallobject}

\begin{figure*}[hbt!]
\centering
\includegraphics[width=\textwidth]{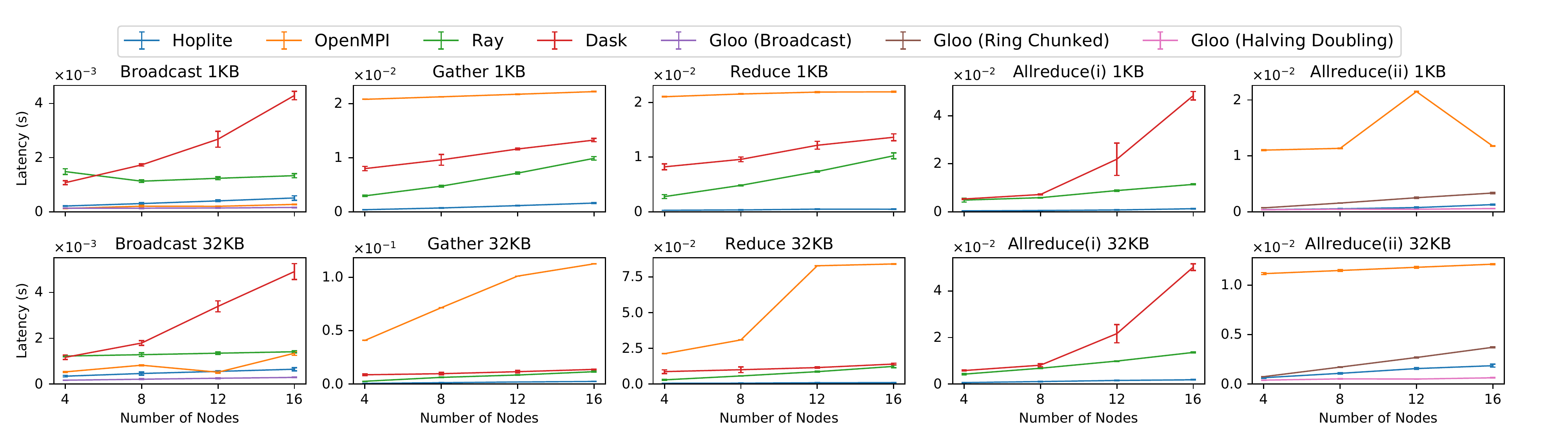}
\caption{\normalfont \small Latency comparison of \codename, OpenMPI, Ray, Dask, and Gloo on standard collective communication primitives (e.g., broadcast, gather, reduce, allreduce) on 1KB and 32KB objects. To show the results more clearly, we split the results of Allreduce into two groups: group (i) includes \codename, Ray, and Dask, and group (ii) includes \codename, OpenMPI, and two different allreduce algorithms in Gloo.}
\label{fig:additional_microbenchmark}
\end{figure*}

We present the microbenchmarks for multiple collective communication primitives for small objects (1KB, 32KB) in Figure~\ref{fig:additional_microbenchmark}. Note that \codename stores object contents in object directory service for objects smaller than 64\,KB (\autoref{sec:directoryservice}), so there is no collective communication for \codename. Again, we compare with Ray, Dask, OpenMPI, and Gloo. We do not compare with Horovod for the same reason that Horovord has three backends: OpenMPI, Gloo, and NCCL. We have already compared with OpenMPI and Gloo. NCCL is for GPU, and \codename currently does not support GPU.

\codename is the best or close to the best among all these alternatives. Gloo has the best performance for broadcast and allreduce. \codename is more efficient than Ray, and Dask because \codename uses stores the object data directly in object directory service.

\section{Ablation Study on Reduce Tree Degree}
\label{sec:ablation}

\begin{figure*}[hbt!]
\centering
\includegraphics[width=\textwidth]{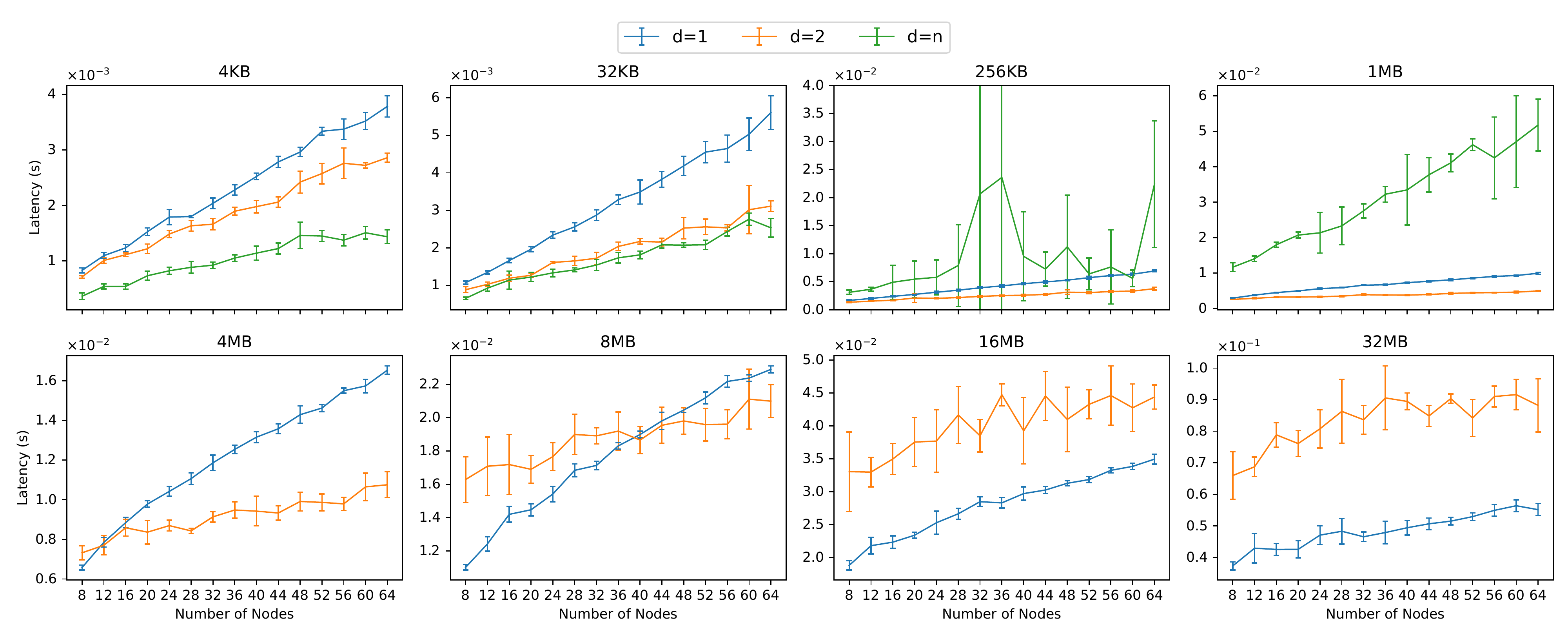}
\caption{\normalfont \small Ablation study of reduce latency on the reduce tree degree $d$ with different object size and number of participants.}
\label{fig:ablation}
\end{figure*}

Here we study the choice of $d$ in the AWS EC2 setting (\autoref{sec:eval}). The best choice of $d$ depends on network characteristics, the size of the object to reduce, and the number of participants. We compare three choices of $d$: 1 (a single chain), 2 (a binary tree), and $n$ (a root connects everyone else). The results are in \autoref{fig:ablation}. As expected from our analysis in (\autoref{sec:coordination}), when the object size is small, $d=n$ is the best because the main bottleneck is the network latency. When the object size is medium (256KB, 1MB), $d=n$ becomes unstable for reduce. We suspect that this is due to incast or due to gRPC characteristics. When object size is 4MB or 8MB, we need to choose between $d=1$ and $d=2$ based on the number of participants. This is because both network latency and network throughput can be a bottleneck in tree reduce. When object size is 16MB or larger, we choose $d=1$ to mitigate the throughput bottleneck in reduce.

\end{document}